\documentclass[12pt]{article}
\usepackage{ametsoc}
\usepackage{epsfig}
\def\diag{\mathop{\rm diag}\nolimits}

\newcommand{\argmin}{\operatornamewithlimits{argmin}}
\newcommand{\vc}[1]{\mathbf{#1}}
\newcommand{\vg}[1]{\mbox{\boldmath{$#1$}}}
\newcommand{\mt}[1]{\textbf{\textsf{{#1}}}}

\newcommand{\myabstract}{
Four-dimensional variational (4D-Var) data assimilation 
on a seasonal-to-interdecadal time-scale
under the existence of unstable modes  
can be viewed as an optimization problem of
synchronized, coupled chaotic systems.
The problem is 
tackled
by 
adjusting initial conditions to bring all stable modes 
closer to observations
and by using a continuous guide to direct unstable modes toward 
a reference time series.
This interpretation provides a consistent and effective 
procedure for solving problems of long-term state estimation.
By applying this approach to an ocean general circulation model with a parameterized
 vertical diffusion procedure, it is demonstrated that 
tangent linear and adjoint models in this framework
should have no unstable modes
and hence be suitable for tracking persistent signals.
This methodology is widely applicable to 
extend the assimilation period in 4D-Var.}
\begin{document}
%
%
\title{\textbf{\large{A framework for interpreting regularized state estimation}}}
%
%
\author{\textsc{Nozomi Sugiura}
				\thanks{\textit{Corresponding author address:} 
				Nozomi Sugiura, 
				Research Institute for Global Change,
				Japan Agency for Marine-Earth Science and Technology,
				2-15 Natsushima-cho, Yokosuka 237-0061, Japan.
				\newline{E-mail:
				nsugiura@jamstec.go.jp}}\quad\textsc{and
				Shuhei Masuda}\\
\textit{\footnotesize{Research Institute for Global Change, JAMSTEC, Yokosuka, Japan}}
\and
\centerline{\textsc{Yosuke Fujii and Masafumi Kamachi}}\\
\centerline{\textit{\footnotesize{Meteorological Research Institute, Tsukuba, Japan}}}
\and 
\centerline{\textsc{Yoichi Ishikawa and Toshiyuki Awaji}}\\
\centerline{\textit{\footnotesize{Data Research Center for Marine-Earth Sciences, JAMSTEC, Yokohama, Japan}}}
}
%
\ifthenelse{\boolean{dc}}
{
\twocolumn[
\begin{@twocolumnfalse}
\amstitle

\begin{center}
\begin{minipage}{13.0cm}
\begin{abstract}
	\myabstract
	\newline
	\begin{center}
		\rule{38mm}{0.2mm}
	\end{center}
\end{abstract}
\end{minipage}
\end{center}
\end{@twocolumnfalse}
]
}
{
\amstitle
\begin{abstract}
\myabstract
\end{abstract}
\newpage
}
%
\section{Introduction}

The 4-dimensional variational (hereafter, 4D-Var) data assimilation 
      method has the advantage that 
it provides a model time-trajectory fit to observations and hence can  create
      a dynamically self-consistent data set.
In particular, it enables estimation of the long-term ocean state,
which is desirable to better 
understand climate change in conjunction with seasonal-to-interdecadal
variation \citep[e.g.,][]{stammer2002,wunsch2007practical,kohl2008decadal,masuda2010simulated}.

The 4D-Var method solves a least-squares problem 
by using gradient information derived from tangent linear and adjoint integrations.
The tangent linear operator is defined as the Jacobian
      matrix of a model's nonlinear propagation operator, and the adjoint
      operator is defined as its transpose.
Despite the existence of these clear mathematical definitions, 
tangent linear and adjoint models (hereafter, linear models) 
that are algebraically approximate
are often used
in practical oceanic or atmospheric applications of 
4D-Var data assimilation.
These inexact linear models are favored not because of
the difficulties in exact differentiation of a complex ocean
general circulation model (OGCM),
but because 
algebraically exact linear models sometimes generate undesirably strong
sensitivities, 
 which give rise to inappropriate gradient information for 
optimizations on a seasonal-to-interdecadal time scale
\citep[e.g.,][]{lea2002sensitivity,hoteit2005treating,kohl2008adjoint}.
Apart from 
the possibility of numerical instabilities,
which cannot be totally ruled out in some applications \citep[e.g.,][]{zhu2000role},
these strong sensitivities have a dynamical origin associated with 
instabilities arising from rapidly growing perturbations that are not
necessarily relevant to the phenomena of interest
\citep{buizza1994sensitivity}.
Previous studies have used some modified linear models 
to address this difficulty and have achieved successful optimization
\citep{hoteit2005treating,sugiura2008development,mazloff2010eddy},
but the rationale for their use is still unclear.
Thus, a systematic procedure for defining the optimization
problem and estimating the long-term ocean state remains elusive.

It is well-known in nonlinear dynamical system research
that a chaotic system can be stabilized by being coupled with an external system
through the mechanism of chaos synchronization
\citep{pecora1990synchronization,pyragas1993predictable}.
Here ``coupling'' refers to a connection between 
two similar dynamical systems in phase space, not to a physical coupling
of air and sea.
A similar concept has long been utilized
in linear control theory.
The Luenberger observer \citep{luenberger1964observing} is a modeled system 
that mimics the truth, and the time evolution of the observer is
described by 
the sum of 
the time-stepping operator of the true system 
and a coupling term, which acts as an attractive force to the truth along some of the
degrees of freedom possessed by the original state.
If the error between the two systems goes to zero, the observer
is synchronized with the truth, or follows exactly the same orbit as the
truth. This condition is called the ``observability condition,''
and it also is characterized by negative conditional Lyapunov exponents for all modes
existing in the observer system.
\citet{so1994observing} extended this concept to nonlinear systems and
established the necessary conditions for the coupling term in a nonlinear
observer to achieve synchronization between the truth and the model.
These conditions are that the variational equation for the observer should be Lyapunov-stable and 
that the image of the Jacobi matrix of the coupling term should remain in
unstable subspace so as not to contaminate the stable modes of the original
system. 
Later, \citet{abarbanel2010data} defined and solved a variational problem with a cost
function that
measures how well the synchronization is achieved between the truth and the nonlinear observer.%

In this paper, we examine these issues from the perspective of 
regularization, that is, 
the smoothing of the cost function surface of
\citet{abarbanel2010data},
by assuming synchronized coupled chaotic systems.
We then 
discuss how to construct appropriate linear models for
an ocean state estimation
that uses an
assimilation window longer than the characteristic period of the fastest growing mode.
We also 
introduce some regularization procedures 
into the vertical diffusion scheme of a realistic OGCM
for practical use.

\section{Formulation}
\subsection{Problem setting}
Data assimilation can be viewed as a problem of synchronizing
two dynamical systems, one representing {\it truth} and the other
representing a {\it model} \citep{yang2006data}.
The former is often called the {\it master system}, whereas the latter is called
the {\it slave system}, because
the flow of information is unidirectional 
from the former to the
latter \citep{duane2006synchronicity}.
%

For instance, \citet{abarbanel2010data} 
achieved a synchronization between two such systems
by introducing
a restoring term to the model equation
that incorporates
as many independent pieces of observational information, or {\it truth}, as 
there are distinct unstable directions in phase space
along which instabilities occur in the synchronization manifold.
(See Appendix A for details.)
The most important implication of their approach is that
to have a stable solution procedure for data assimilation,
the model should provide some restoring terms, or coupling terms,
toward a {\it truth}
to suppress the instabilities that may occur in some modes.
In the case of ocean data assimilation,
although the model 
that describes the ocean dynamical system 
exhibits an unstable nature, 
we usually cannot provide 
enough observations 
to 
assign to all of the unstable directions.
We rather focus on chaos synchronization between the two systems,
or the stabilization of the slave system 
by the continuous guidance
of some master system.
The concept of synchronization is still applicable even if
{\it truth} is replaced by some other external system with an 
evolution law similar to that of the {\it model}.
That is, 
although in geophysical applications we usually do not have enough
observational information 
to restore all distinct unstable directions,
we are still able to treat the information from {\it truth} as hidden variables
and assign them
tentative values, 
provided by a model integration from 
an updated guess of the initial condition
(called a ``reference time series'' hereafter).
Under the assumption that this kind of additional external
information is available, 
a regularized data assimilation problem,
of minimizing a cost function (Eq.\,(\ref{4dvar1})) 
subject to constraints (Eqs.\,(\ref{4dvar2}) and (\ref{4dvar3})),
is defined as follows. 

The original 4D-Var problem in an incremental formulation
\citep[e.g.,][]{courtier1994strategy,lawless2005investigation,tremolet2007incremental}
is defined as the problem of finding
an optimal initial condition,
under the constraint of observations,
that  approximately minimizes the cost function
    \begin{eqnarray}
     \mathcal{J}(\vg{\psi};\vg{\theta})
 &=&
 \frac12
 \left(
 \vg{\theta}+\vg{\psi}-\vg{\theta}_b
 \right)^T
 \mt{B}^{-1}
 \left(
 \vg{\theta}+\vg{\psi}-\vg{\theta}_b
 \right)\nonumber\\
 &&+
 \frac12
 \sum_{n=1}^N 
 \left. 
 \left( \vc{H}(\vc{x})-\vc{x}^{obs} \right)^T
 \mt{R}^{-1}
 \left( \vc{H}(\vc{x})-\vc{x}^{obs} \right) 
 \right|_{t=t_n} 
 ,\label{4dvar1}
 \end{eqnarray}
subject to 
   \begin{eqnarray}
\dot{\vc{x}} &=& \vc{f}(\vc{x}),\quad
 \vc{x}(0)=\vg{\theta}+\vg{\psi},\label{4dvar3orig}
   \end{eqnarray}
with the reference time series for gradient calculation defined as
   \begin{eqnarray}
\dot{\vc{y}} &=& \vc{f}(\vc{y}),\quad
 \vc{y}(0)=\vg{\theta}, \label{4dvar2}
\end{eqnarray}
where 
$n$ is the time index,
$\mt{B}$ is the background error covariance matrix,
$\mt{R}$ is the observational error covariance matrix,
$\vg{\theta}$ is the initial state of the reference system,
$\vg{\psi}$ is the difference between the initial states of 
the estimated systems and the reference system,
$\vg{\theta}_b$ is a firstguess for the initial state of the reference system,
$\vc{x}$ denotes the estimated time series integrated from the initial state
$\vg{\theta}+\vg{\psi}$,
$\vc{y}$ denotes the reference time series integrated from the initial state
$\vg{\theta}$,
and $\vc{x}^{obs}$ denotes the observations.
The optimization problem will be approximately solved for the control
variable $\vg{\psi}$, 
a procedure that is called the inner loop.
Note that the inner loop only concerns  the incremental state $\vc{v}=\vc{x}-\vc{y}$.
Although $\vg{\theta}$, or $\vc{y}$, is usually updated as well using the information from the
optimized value of $\vg{\psi}$, 
which is called the outer loop of incremental 4D-Var,
we concentrate here on the optimization
of $\vg{\psi}$, or $\vc{v}$, in the inner loop.%

For simplicity, we assume in Eq.\,(\ref{4dvar1}) 
that
$\vg{\theta}=\vg{\theta}_b$,
$\mt{B}=\sigma_b^2 \mt{I}$, 
$\mt{R}=\sigma_o^2\mt{I}$, 
$\vc{H}=\mathcal{I}$, the identity operator, 
and all observations $\vc{x}^{obs}$ are located at the end of the
assimilation window $(t=t_N)$.
Accordingly, 
we can consider that
the model operator $\vc{M}$ assigns to every initial state
$\vc{x}(0)$ a final state $\vc{x}(t_N)$.
In this setting, the incremental 4D-Var cost function in quadratic form 
is written as
    \begin{eqnarray}
     \mathcal{J}\left(
		 \vg{\psi}
       \right)
      &=&
      \frac1{2\sigma_b^2} \vg{\psi}^T \vg{\psi}
      +
      \frac1{2\sigma_o^2}
      \left( \mt{M}\vg{\psi}-\vc{d} \right)^T 
      \left( \mt{M}\vg{\psi}-\vc{d} \right),\label{eq:simple_j}\\
     \vc{d}&=& \vc{x}^{obs}-\vc{M}(\vg{\theta}_b),
    \end{eqnarray}
where 
$\mt{M}$ is the derivative of the model operator $\vc{M}$
with respect to the initial condition $\vg{\psi}$ of the 
incremental state $\vc{v}$.
It is convenient to express
$\mt{M}$
in the form of a singular value decomposition \citep{johnson2006singular}: 
    \begin{eqnarray}
     \mt{M}&=& \sum_j \vc{U}_j \sigma_j \vc{V}_j^T,\\
     \sigma_1 > \sigma_2 & > & \cdots  >  \sigma_{j_s-1}  > 1  \ge 
      \sigma_{j_s} > \cdots  >  \sigma_{r}\sim 0,
    \end{eqnarray}
where $r$ is the rank of $\mt{M}$, $j_s$ indicates the first stable mode, 
and, for each mode $j$, $\vc{V}_j$ is the right singular vector,
$\vc{U}_j$ is the left singular vector,
and $\sigma_j$ is the singular value, or
the growth rate during the assimilation
window.
Each set of singular vectors is orthonormal and can be regarded as 
a finite-interval counterpart of the
Lyapunov vectors \citep{legras1996guide}.
Using this decomposition, the gradient
used for the inner loop optimization is expressed as
    \begin{eqnarray}
     \nabla \mathcal{J}
      &=&
      \frac1{\sigma_b^2} \vg{\psi}
      +
      \frac1{\sigma_o^2}
      \mt{M}^T 
      \left( \mt{M}\vg{\psi}-\vc{d} \right)\\
    &=& \sigma_o^{-2} \sum_j \left[  
		\left(  \mu^2+\sigma_j^2
		\right) \left( \vc{V}_j^T \vg{\psi} \right)
-\sigma_j \vc{U}_j^T \vc{d}
\right] \vc{V}_j,
    \end{eqnarray}
where $\mu=\sigma_o/\sigma_b$.
If the fastest growth rate $\sigma_1$ becomes many orders of magnitude larger than $1$ due to the
extension of the assimilation window beyond the predictability,
then the cost function $\mathcal{J}$ becomes so sensitive to the initial condition
$\vg{\psi}$, or the magnitude of $\nabla \mathcal{J}$ becomes so large,
that any gradient-based solution method can hardly solve the
stationary problem $\nabla \mathcal{J}=0$, or more precisely, 
almost infinite precision is required 
for an initial condition to solve it.
This situation corresponds to 
what \citet{abarbanel2010data} called
the irregularity of the cost function surface.

Even in that situation, we can still solve the problem in a reduced
control space spanned by the stable singular vectors $\vc{V}_{j_s},\vc{V}_{j_s+1},\cdots$.
In fact, using a projection operator $\mathcal{P}_s=\sum_{j \ge j_s} \vc{V}_j \vc{V}_j^T$,
we can define a new cost function: 
    \begin{eqnarray}
     \tilde{\mathcal{J}}\left(
\vg{\psi}
\right)
&\equiv&
     \mathcal{J}\left(
\sum_{j \ge j_s} \vc{V}_j \vc{V}_j^T \vg{\psi}
\right)\nonumber\\
      &=&
      \frac1{2\sigma_b^2} \sum_{j \ge j_s} \left( \vc{V}_j^T \vg{\psi} \right)^2
      +
      \frac1{2\sigma_o^2}
      \left( \tilde{\mt{M}}\vg{\psi}-\vc{d} \right)^T 
      \left( \tilde{\mt{M}}\vg{\psi}-\vc{d} \right),\label{eq:reduced_J}\\
     \tilde{\mt{M}}&=& \sum_{j \ge j_s} \vc{U}_j \sigma_j \vc{V}_j^T,
    \end{eqnarray}
of which the gradient
    \begin{eqnarray}
     \nabla \tilde{\mathcal{J}}
    &=& \sigma_o^{-2} \sum_{j\ge j_s} \left[  
		\left(  \mu^2+\sigma_j^2
		\right) \left( \vc{V}_j^T \vg{\psi} \right)
-\sigma_j \vc{U}_j^T \vc{d}
\right] \vc{V}_j
\label{eq:reduced_dJ}
    \end{eqnarray}
is now appropriately 
derived using the growth rates $\sigma_j \le 1~(j \ge j_s)$.
The point is that the definition of the new cost function $\tilde{\mathcal{J}}$ is based on identifying
the stable singular vectors.
If the image of the projection operator $\mathcal{P}_s$ contains 
even a very small fraction of unstable modes, 
then it will grow exponentially, and thus the
reduced treatment will fail.

In practice, a complete set of stable singular vectors
cannot be perfectly prescribed in an ocean data assimilation,
and thus we should use instead a modified model $\hat{\vc{M}}$ 
which has a linearized dynamics $\hat{\mt{M}}$ similar to $\mt{M}$ but
exhibits a stable evolution. 
If each growth rate $\sigma_j$ is assumed to be modified to
$\hat{\sigma}_j \le 1$ in the linearized dynamics $\hat{\mt{M}}$, then we can define another new cost
function $\hat{\mathcal{J}}$ as
    \begin{eqnarray}
     \hat{\mathcal{J}}\left(
		       \vg{\psi}
       \right)
     &=&
      \frac1{2\sigma_b^2} \vg{\psi}^T \vg{\psi}
      +
      \frac1{2\sigma_o^2}
      \left( \hat{\mt{M}}\vg{\psi}-\vc{d} \right)^T 
      \left( \hat{\mt{M}}\vg{\psi}-\vc{d} \right),\label{eq:reduced2_J}\\
     \hat{\mt{M}}&=& \sum_{j} \vc{U}_j \hat{\sigma}_j \vc{V}_j^T,
    \end{eqnarray}
of which the gradient 
    \begin{eqnarray}
     \nabla \hat{\mathcal{J}}
    &=& \sigma_o^{-2} \sum_{j} \left[  
		\left(  \mu^2+\hat{\sigma}_j^2
		\right) \left( \vc{V}_j^T \vg{\psi} \right)
-\hat{\sigma}_j \vc{U}_j^T \vc{d}
\right] \vc{V}_j
\label{eq:reduced2_dJ}
    \end{eqnarray}
is also appropriately derived using the growth rates $\hat{\sigma}_j \le 1~(j=1,2,\cdots)$.
Note that this cost function $\hat{\mathcal{J}}$ includes $\tilde{\mathcal{J}}$
as a special case with $\hat{\sigma_j}=0 ~(j < j_s)$ and
$\hat{\sigma_j}=\sigma_j  ~(j \ge j_s)$, except for a slight difference in
the background term.
To prepare a model $\hat{\vc{M}}$ appropriate for $\hat{\mathcal{J}}$, 
we introduce into the nonlinear model a restoring term,
similar to that of \citet{abarbanel2010data},
that is intended to suppress the
unstable modes that may arise at every moment of the integration of its
linear models.
In what follows, the mapping
$\hat{\vc{M}}: \vc{x}(0) \longmapsto \vc{x}(t_N)$
is represented in the form of a differential equation for $\vc{x}$
(e.g., Eq.\,(\ref{4dvar3})).
%

Now we are ready to specify how
regularized 4D-Var 
makes tractable optimization problems
that are insoluble by the original 4D-Var.
The modification of the constraints on the cost function (Eq.\,(\ref{4dvar1})) is made
by assuming the following form of the model evolution
instead of Eq.\,(\ref{4dvar3orig}):
   \begin{eqnarray}
 \dot{\vc{x}} &=& \vc{f}(\vc{x}) + \epsilon \vc{g}(\vc{x},\vc{y}),\quad
  \vc{x}(0)=\vg{\theta}+\vg{\psi},\label{4dvar3}
   \end{eqnarray}
where $\vc{g}$ is an anti-symmetric function 
and $\epsilon$ is a coupling intensity.
In this formulation, we have 
a master system $\vc{y}$
described by a nonlinear model 
integrated from a
known initial state $\vg{\theta}$,
and a slave system $\vc{x}$
described by a model, using the same equation
but with a coupling term between them, 
that is integrated from an
estimated initial state $\vg{\theta}+\vg{\psi}$.
The slave
system is attracted to the master system in phase space 
through the coupling term
(Figure \ref{f1}).
%
As shown in Appendix B,
the
variational equation (the tangent linear equation
that describes sensitivities) 
along the transverse direction $\vc{v}=\vc{x}-\vc{y}$
on the synchronization manifold $\vc{x}=\vc{y}$ is
   \begin{eqnarray}
    \dot{\delta \vc{v}} =& 
     \left( \mt{Df} + \epsilon \mt{D}_1\mt{g} \right) \delta
     \vc{v},
     \label{eq:variational}
   \end{eqnarray}
which is decoupled from the sensitivities of the master system.%

\begin{figure}
\includegraphics[width=0.9\linewidth]{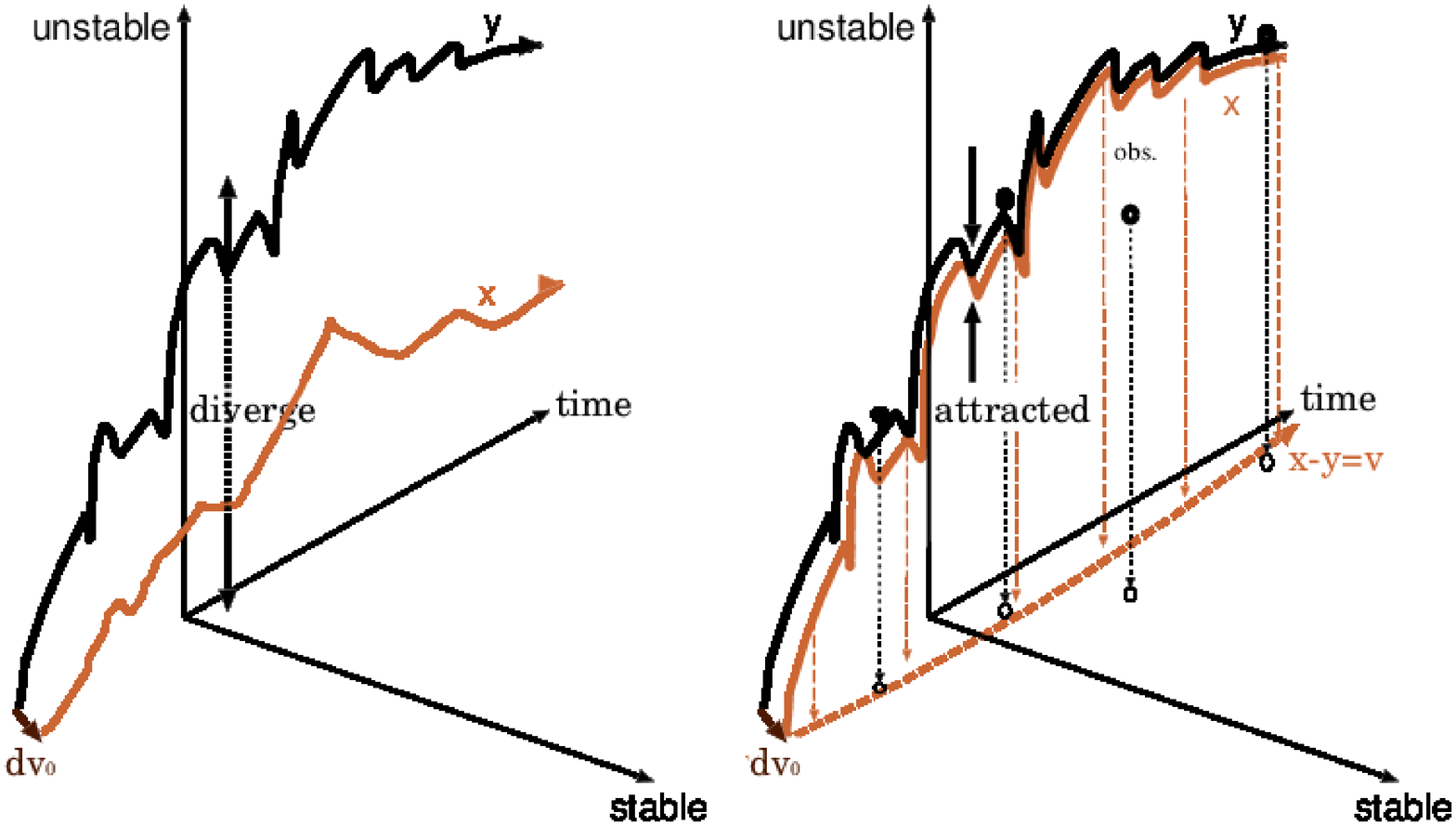}
\caption{
Schematic views of two chaotic systems that are close at the
 starting period. 
Two degrees of freedom for each system are shown as axes, one of which
 is linearly stable, whereas the other is unstable.
Uncoupled systems diverge with time (left), whereas
 coupled systems synchronize (right) because of the attractive force
 between the two unstable variables.
}
\label{f1}
\end{figure}

The largest conditional Lyapunov exponent for the variational equation (\ref{eq:variational})
can be reduced by properly defining the coupling function $\vc{g}$
\citep[e.g.,][]{pyragas1993predictable}. 
To that end, $\vc{g}$ should be designed as an attractive force
so that $\mt{D}_1\mt{g}$ 
provides a damping effect on 
the eigenspaces with positive Lyapunov exponents of $\mt{Df}$, 
which correspond to unstable directions in phase space.

We can characterize the regularized 4D-Var method in comparison with the other
methods mentioned as follows.
The method of \citet{abarbanel2010data} 
adjusts initial conditions and restoring coefficients to bring
all unstable but stabilized modes closer to observations 
and use a continuous guide 
to bring the model's unstable modes toward observational data.
The original 4D-Var method
adjusts initial conditions to bring all modes 
closer to observations.
The regularized 4D-Var method
adjusts initial conditions to bring all stable modes 
closer to observations
and uses a continuous guide to direct the unstable modes toward 
the reference time series.
The similarity between this method and that of \citet{abarbanel2010data} lies in 
the use of regularization, which
makes all modes stable in the model with some guidance by external systems 
(the reference time series and observations, respectively).%
As is shown below, this framework offers a unified interpretation of the
existing regularization methods with modified adjoint models \citep[e.g.,][]{hoteit2005treating}.
See Appendix C for a discrete-time version of the regularized 4D-Var problem
and the details of the solution algorithm for practical use. 

Here we show an application of this kind of regularization 
to linear models for data assimilation,
where coupled master-slave systems are implicitly assumed. 
Horizontal eddy activities are a major cause of
instability in the ocean, resulting in large sensitivities 
of linear models.
\citet{hoteit2005treating} suppressed these large sensitivities by introducing a 
horizontal diffusion coefficient 
and a horizontal viscosity coefficient 
(horizontal mixing coefficients hereafter)
larger than the ones in the forward model.
In the context of regularization, we interpret their method as follows. 
The object to which data assimilation is applied is defined by
substituting $\vc{f}(\vc{x})=\vc{f}_0(\vc{x})+\mt{K}\nabla^2 \vc{x}$ and 
$\vc{g}(\vc{x},\vc{y})=-\mt{K} \nabla^2 (\vc{y}-\vc{x})$
into Eqs.\,(\ref{4dvar2}), (\ref{4dvar3}), and (\ref{eq:variational}).
Because the coupling term is related to a term in the original 
nonlinear model operator $\vc{f}$, here we set the relevant term apart from
the rest of the terms $\vc{f}_0$.
The nonlinear and variational equations for this case are
   \begin{eqnarray}
\dot{\vc{y}} &=& \vc{f}_0(\vc{y}) + \mt{K} \nabla^2 \vc{y}, \label{hdiff_sens1}\\
\dot{\vc{x}} &=& \vc{f}_0(\vc{x}) + \mt{K} \nabla^2 \vc{x} 
- \epsilon \mt{K} \nabla^2 (\vc{y}-\vc{x}),\label{hdiff_sens2}
\end{eqnarray}
and
   \begin{eqnarray}
\dot{\delta \vc{v}} &=& \left[\mt{Df}_0+ (1+\epsilon) \mt{K} \nabla^2 \right] \delta \vc{v},\label{hdiff_sens3} 
   \end{eqnarray}
where $\vc{f}_0$ is 
the model operator excluding the mixing terms, $\mt{K} \geq 0$ 
represents the horizontal mixing coefficients
(constant tensor),
$\nabla^2$ denotes the horizontal Laplacian, 
and $\epsilon$ is the coupling intensity.

Thus, the variational equation is the same as the original tangent
linear equation except that the mixing tensor $\mt{K}$ is
enhanced to $(1+\epsilon)\mt{K}$.
That is, the enhanced horizontal mixing coefficients are
explained by
coupling with a master system.

\subsection{Role of the coupling term}
For the regularization to work properly,
an efficient coupling function 
must meet three criteria:
\begin{enumerate}
\item For the separation of the sensitivities, $\vc{g}$ must be
      antisymmetric. \label{cond_a}
\item For the stability of the $\vc{v}$-system, the evolution governed by 
$\dot{\delta \vc{v}}=(\mt{D}\mt{f}+\epsilon
      \mt{D}_1\mt{g})\delta\vc{v}$ must have no positive Lyapunov
      exponents. \label{cond_b}
\item For the similarity between the sensitivities of the $\vc{v}$-system and $\vc{y}$-system,
the image of $\mt{D}_1\mt{g}$
 should lie mainly in an unstable subspace
in order not to contaminate too much the evolution of
      Lyapunov vectors in the originally stable subspace. 
In other words,
the stable subspace of the $\vc{v}$-system should not
lose its original features in the stable subspace of the $\vc{y}$-system. 
\label{cond_c}
\end{enumerate}

One of the ideal coupling functions that fully satisfies conditions \ref{cond_a} and \ref{cond_c} is 
found in the master and slave systems of \citet{abarbanel2010data},
given here as Eqs.\,(\ref{master_Ab}), (\ref{slave_Ab}), and
(\ref{coupling_Ab}) in Appendix A.
The variational equation for the difference between the $\vc{x}$- and $\vc{y}$-systems is
\begin{eqnarray}
\dot{\delta \vc{v}} &=& 
\left(
\mt{Df}
-
\begin{bmatrix}
 \mt{C}(t)& \\
 &\mt{O}_{M-M_u}
\end{bmatrix}
\right)
\delta \vc{v}.
\label{eq:ideal_coupling}
\end{eqnarray}
\citet{abarbanel2010data} also controlled the function $\mt{C}(t)$ 
to meet condition \ref{cond_b} by a variational method
(Eq.\,(\ref{cost_abarbanel}) in Appendix A).
When it is  difficult to separate the unstable directions
cleanly, as they did, such as in the case of a high dimensional system,
we should find an acceptable candidate for $\vc{g}$ 
that reasonably satisfies the above conditions.%

Below we discuss the role of the coupling term $\epsilon \vc{g}$ and how
it should be efficiently defined in practical applications, 
taking the case of horizontal mixing coefficients
as an example.
We assume here, for simplicity, that $\delta \vc{v}$ corresponds to a
function of 1-dimensional space $s$ and time $t$, $\delta v(s,t)$,
which is expressed as a Fourier series on a finite spatial interval $[0,L]$,
 \begin{eqnarray}
  \delta v(s,t) &=& \sum_{j=0}^{\infty} \delta v_j(s,t),\\
  \delta v_j(s,t) &\equiv& a_j^0(t) \cos{\left(\frac{2 \pi j}{L}s \right)}+a_j^1(t) \sin{\left(\frac{2 \pi j}{L}s \right)}.
\end{eqnarray}
The action of $\mt{Df}_0$ on the $j$-th mode is given by a growth rate
$\lambda_j$,
 \begin{equation}
\mt{Df}_0\delta v_j=\lambda_j \delta v_j.
\end{equation}
Substituting the above expressions into Eq.\,(\ref{hdiff_sens3}), we get
 \begin{eqnarray}
\frac{\dot{a}_j^l}{a_j^l} &=& \lambda_j-(1+\epsilon)K\left( \frac{2\pi j}{L}
					 \right)^2, \quad l=0,1, \label{eq_for_a}
 \end{eqnarray}
where the mixing coefficients are treated as a scalar constant $K$ for simplicity.
The action  of the coupling term $\mt{D}_1\mt{g}$
is represented here by the terms 
$-K\left( 2\pi j/L \right)^2$, which constitute
an attractive force to the master system along the high-wavenumber
subspace in substitution for the unstable subspace;
that is, the coupling term attracts the two states
along high-wavenumber horizontal variability.
The larger the wavenumber is, the
more strongly $\vc{x}$ is attracted to $\vc{y}$.

The solutions of Eq.\,(\ref{eq_for_a}) are
 \begin{eqnarray}
  a_j^l(t) &=& a_j^l(0)\exp{\left\{\left[ \lambda_j-(1+\epsilon)K\left( \frac{2\pi j}{L}
					 \right)^2 \right]
  t\right\}}, \quad l=0,1. \label{sol_a}
 \end{eqnarray}
If we have $\epsilon=0$ ($\epsilon>0$) in Eq.\,(\ref{sol_a}), then it describes the error
growth of the $j$-th Fourier mode of the $\vc{y}$-system (the $\vc{v}$-system). 
The condition for stabilizing the $\vc{v}$-system (condition \ref{cond_b}) is 
that 
for every $j$ that satisfies 
$\lambda_j-K\left( \frac{2\pi j}{L}\right)^2 >0$
the following must be satisfied:
 \begin{equation}
\lambda_j-(1+\epsilon)K\left( \frac{2\pi j}{L}\right)^2 \le0.\label{stab_eps}
\end{equation}
The application of $\epsilon>0$ also has an influence on the originally stable modes which satisfy
$\lambda_j-K\left( \frac{2\pi j}{L}\right)^2 <0$.
To minimize the impact of the coupling term on
the stable subspace of the original system (condition \ref{cond_c}), we should choose the smallest
possible $\epsilon$ that satisfies the stability condition of Eq.\,(\ref{stab_eps}).
Note that 
we are not directly assigning a value to $\epsilon$ for any actual application,
because we made here several assumptions about the system configuration and
parameters.%

In terms of the analysis field, 
the effect of the coupling term on the solution is assessed as follows.
For the original 4D-Var problem, 
the minimizer of the cost function
(Eq.\,(\ref{eq:simple_j})) is formally written with singular vectors
as
    \begin{eqnarray}
     \vg{\psi}_a &=& 
      \sum_{j}
      \frac{{\sigma}_j^2}{\mu^2+{\sigma}_j^2} \frac{\vc{U}_j^T
      \vc{d}}{{\sigma}_j} \vc{V}_j,
    \end{eqnarray}
where $\mu = \sigma_o / \sigma_b$ \citep{johnson2006singular}.
Note that this solution is not available by any gradient method when the
assimilation window is much longer than the predictability of the fastest
mode.
%
With the ideal coupling function (Eq.\,(\ref{eq:ideal_coupling})), we 
now can write the
minimizer of the regularized cost function (Eq.\,(\ref{eq:reduced2_J})) as
    \begin{eqnarray}
     \hat{\vg{\psi}}_a &=& 
      \sum_{j \ge j_s}
      \frac{\sigma_j^2}{\mu^2+\sigma_j^2} \frac{\vc{U}_j^T
      \vc{d}}{\sigma_j} \vc{V}_j
      +
      \sum_{j < j_s}
      \frac{\hat{\sigma}_j^2}{\mu^2+\hat{\sigma}_j^2} \frac{\vc{U}_j^T
      \vc{d}}{\hat{\sigma}_j} \vc{V}_j,
    \end{eqnarray}
where $\hat{\sigma}_j ~(j < j_s)$ 
correspond to originally unstable modes
that are stabilized by the additional coefficients
$-\mt{C}(t)$ in Eq.\,(\ref{eq:ideal_coupling}).
In this solution, all the modes in originally stable subspace ($j \ge
j_s$) remain unchanged.
Usually, these modes are also affected by the coupling term,
and the solution is changed into
    \begin{eqnarray}
     \hat{\vg{\psi}}_a &=& 
      \sum_{j}
      \frac{\hat{\sigma}_j^2}{\mu^2+\hat{\sigma}_j^2} \frac{\vc{U}_j^T
      \vc{d}}{\hat{\sigma}_j} \vc{V}_j,
    \end{eqnarray}
where  $\hat{\sigma}_j \le 1$.
The rate of deformation
on the $j$-th mode of the analysis increment at time $t_N$,
caused by the regularization,
is estimated 
as
    \begin{eqnarray}
\frac{
\left(  \vc{U}_j,
     \hat{\mt{M}}\hat{\vg{\psi}}_a
-
     \mt{M}\vg{\psi}_a
\right)
}
{
\left(  \vc{U}_j,
     \mt{M}\vg{\psi}_a
\right)
}
 &=& 
\frac{
\left( \frac{\hat{\sigma}_j}{\sigma_j} \right)^2 - 1
}
{
1 + \left(\frac{\hat{\sigma}_j}{\mu} \right)^2 
}.
\label{DHMphi}
    \end{eqnarray}
%
In the case of enhanced horizontal mixing (Eq.\,(\ref{sol_a})), 
the rate of deformation is obtained by substituting
      \begin{eqnarray}
       \frac{\hat{\sigma}_j}{\sigma_j} &=&  \exp{\left[ -\epsilon K\left( \frac{2\pi j}{L}
					 \right)^2 
  t_N\right]}.
\label{dSigma}
      \end{eqnarray}
Given that $\hat{\sigma}_j$ should not be much larger than $\mu$,
Eqs.\,(\ref{DHMphi}) and (\ref{dSigma})
represent the fact that the deformation in the analysis increment
has a property of higher wave-number modes, possibly including
stable modes, being damped more strongly.%

Thus, 
the most significant effect of the coupling term is that it enables us
to 
solve the optimization problem in an assimilation window
much longer  than the one used in the original 4D-Var method, although
it confines the analysis increment to modes that are stable in the original
dynamics,
possibly causing some deformation in the increment.
Nevertheless, an advantage of this approach 
is that we can extend the assimilation window
with minimal additional computational cost 
other than that needed for the extended integration of the linear models.

\section{Regularization for vertical mixing schemes}\label{bsec_vert}
\subsection{Methodology}\label{subsec2}
In OGCMs, 
the oceanic instability associated with the vertical mixing process
is expressed through the parameterization of mixed-layer dynamics.
The sensitivity arising from the variation in this parameterization is a major
obstacle to deriving effective gradient information in data
assimilation.
In some previous implementations
\citep[e.g.,][]{hoteit2005treating,gebbie2006strategies,kohl2008decadal,sugiura2008development},
this sensitivity was suppressed 
by not taking into account
the variation of vertical diffusion coefficients
or 
by omitting the linearization of the mixed-layer parameterization.
\citet{Zhu2002} achieved a successful optimization 
by strategically ignoring the part of 
the variation of vertical diffusion coefficients 
that was caused by the variation of turbulent kinetic energy.
Otherwise, the linear models of the ocean mixed-layer parameterization
in these implementations
would have caused the strong sensitivities to hide other important gradient information.

The passive treatment of the vertical diffusion coefficients in
the 4D-Var method 
allows an interpretation 
that 
the model constraint is expressed by a slave system 
whose vertical diffusion coefficients 
are attracted to the ones defined in the master system.
The model is constructed 
by substituting 
$\vc{f}(\vc{x})
=\vc{f}_0(\vc{x})+\vc{k}(\vc{x})\circ\nabla_z^2
\vc{x}$ and 
$\vc{g}(\vc{x},\vc{y})=\left(
\vc{k}(\vc{y})-\vc{k}(\vc{x}) \right)
\circ\nabla_z^2(\vc{x}+\vc{y})/2$
into Eqs.\,(\ref{4dvar2}), (\ref{4dvar3}), and (\ref{eq:variational}),
where $\circ$ denotes the entrywise product.
The nonlinear and variational equations
for the case with parameterized vertical diffusion coefficients are then 
   \begin{eqnarray}
    \dot{\vc{y}} &=& \vc{f}_0(\vc{y}) + \vc{k}(\vc{y})
    \circ \nabla_z^2 \vc{y}, \label{vdiff_sens1}\\
    \dot{\vc{x}} &=& \vc{f}_0(\vc{x}) +
    \vc{k}(\vc{x}) \circ\nabla_z^2 \vc{x} 
    + \epsilon \left( \vc{k}(\vc{y})-\vc{k}(\vc{x}) \right)
    \circ\nabla_z^2 
    \vc{u}, \label{vdiff_sens2}\\
    \dot{\delta \vc{v}} &=&  
    \left[\mt{Df}_0+(1-\epsilon) (\nabla_z^2 \vc{u}) \circ\mt{Dk} 
     +\vc{k}(\vc{u}) \circ \nabla_z^2 \right] \delta \vc{v},\label{vdiff_sens3}
   \end{eqnarray}
where $\vc{f}_0$ is the model operator, $\vc{k}(\vc{x}) \geq
0$ is the vertical
diffusion coefficient (vector function of $\vc{x}$),
$\vc{u}\equiv(\vc{x}+\vc{y})/2$, 
$\nabla_z^2$ is the vertical Laplacian,
and $\epsilon$ is the coupling intensity.

Eq.\,(\ref{vdiff_sens3}) means that the variational equation is the same as the original tangent
linear equation except that the functional dependency of the diffusion coefficient $\vc{k}$ is
reduced to $(1-\epsilon)\mt{Dk}$.
The operation $(\nabla_z^2 \vc{u}) \circ \mt{Dk} $ has eigenvalues with 
indefinite sign,
whereas the operation $\vc{k} \circ \nabla_z^2$ has only negative eigenvalues,
the reduction of the former term can make the combined eigenstructure more stable.
In particular, $\epsilon=1$ means that the diffusion coefficient is
completely prescribed, as was done by  \citet{sugiura2008development}, 
and smaller values correspond to a partially prescribed treatment.


%
Here we explain how regularization for vertical mixing schemes
works by assuming a simplified situation.
We assume that $\delta \vc{v}$ corresponds to a
function of space $s$ and time $t$, $\delta v(s,t)$.
The space $s$ does not necessarily mean a vertical 1-dimensional one, but
it can also  have horizontal spans.
Introducing a space-time white noise $\eta$, which mimics the operation $(\nabla_z^2 \vc{u}) \circ \mt{Dk}$,
we can describe the evolution of a mode in the $\vc{v}$-system
(Eq.\,(\ref{vdiff_sens3})) as
\begin{eqnarray}
\dot{\delta v}&=& \left[ \lambda  + (1-\epsilon)\eta+ k \nabla^2 \right] \delta v,
\end{eqnarray}
where $\delta v$ is the function that represents the mode,
$\lambda$ is the growth rate caused by $\mt{Df}_0$,
and the background values of the mixing coefficients are represented by a scalar constant $k$ for simplicity.
This noisy heat equation can exhibit complicated behavior
due to the combination of a multiplicative noise
term $\eta \delta v$ and a spatial correlation term $k\nabla^2 \delta v$.
By applying a logarithmic transformation $w=\log{\left| \delta v \right|}$, we get
       \begin{equation}
	\dot{w}= \lambda
+ (1-\epsilon)\eta
+ k \nabla^2 w
+k(\nabla w)^2, 
	 \end{equation}
which is known as the Karder-Parisi-Zhang (KPZ) equation
\citep{kardar1986dynamic}. 
At least in the spatially 1-dimensional case, it is known that the
nonlinear term $k(\nabla w)^2$ starts to become relevant when
the magnitude of the noise $(1-\epsilon)\eta$ exceeds a critical value
\citep{kapral1994dynamics}. 
Although much is unknown about the behavior in  multi-dimensional cases,
it is reported that the KPZ equation also captures the dynamics of the
logarithmic error growth of a
global weather model \citep{primo2007error}. 
If we assume the existence of a critical value, 
we can control the stability of the system by adjusting the noise level,
 which changes
according to the value of $\epsilon$.
Apparently, $\epsilon=1$ corresponds to a noiseless and stable situation,
but if $\epsilon$ is less than a certain value the system can become
unstable because of the roughness term $(\nabla w)^2$.

This kind of regularization, regarding coefficients with
indefinite signs in a variational equation, might also be applicable to
advection terms in an OGCM.
We describe a possible procedure in Appendix D.

\subsection{Case study}
To demonstrate the effect of regularization 
on the sensitivities 
used in 4D-Var data assimilation into a state-of-the-art OGCM, 
we conducted a comparative study of three
linearized treatments of the vertical diffusion coefficient 
in the master-slave setting described above.
The model we used was based on the Meteorological Research Institute (MRI) Community Ocean Model
\citep{mricom1,mricom2} developed by 
the Japan Meteorological
Agency.
It is a global model with a horizontal
resolution of $1^{\circ}$ longitude and $0.5^{\circ}$ latitude with 51
vertical levels.
The model is integrated 
under a climatological atmospheric forcing through a bulk
parameterization scheme 
and with the mixed-layer closure scheme of \citet{noh1999simulations}.
The model has an algorithmic structure in which
the prognostic variables $\vc{y}$ are integrated using the vertical
diffusion coefficients $\vc{k}$, which are derived from the turbulent kinetic
energy described by the prognostic mixed-layer
dynamics 
for $\vc{y}$. 
Hence, the whole system with regularization 
is compactly described 
by Eqs.\,(\ref{vdiff_sens1}), (\ref{vdiff_sens2}), and (\ref{vdiff_sens3}).
By using a tangent linear system governed by Eq.\,(\ref{vdiff_sens3}),
we compared the effect of regularization under the following three settings.

\begin{itemize}
\item Case 1\label{case1}.
The variation of the vertical diffusion coefficients is derived dynamically from
the linearized version of the prognostic mixed-layer dynamics
($\epsilon=0$ in Eqs.
      (\ref{vdiff_sens2}) and (\ref{vdiff_sens3})).
\item Case 2\label{case2}.
The values of the vertical diffusion coefficients are prescribed by a master system
($\epsilon=1$ in Eqs.
      (\ref{vdiff_sens2}) and (\ref{vdiff_sens3})), as has 
been commonly done 
in previous studies
\cite[e.g.,][]{sugiura2008development}.
\item Case 3\label{case3}.
The vertical diffusion coefficients are partly prescribed, but their
      variation, which is
derived from the linearized version of the prognostic mixed-layer dynamics,
is also partly taken into account
($\epsilon=0.75$ in Eqs.
      (\ref{vdiff_sens2}) and (\ref{vdiff_sens3})).
\end{itemize}

\subsection{Results}
We tested the stability of the linear models by evaluating
the first backward Lyapunov vectors \citep{legras1996guide}, that is,
the modes that have grown the fastest,
and the corresponding Lyapunov exponents.
The $j$-th Lyapunov exponent is defined as
       \begin{eqnarray}
	\lambda_j &=& \lim_{t \to \infty}\frac1t \log{\| \vg{\xi}_j(t) \|},
       \end{eqnarray}
by using $\vg{\xi}_j(t)$, the $j$-th Lyapunov vector at time $t$,
and the norm of the vector, 
	\begin{eqnarray}
	 \| \vg{\xi}_j(t) \|^2 &=& \frac{\sum_m \left( \vg{\xi}_j^m(t) \right)^2 \delta V_m}{\sum_m \delta V_m},
	\end{eqnarray}
where 
$\xi^m_j(t)$ is the $m$-th component of the $j$-th Lyapunov vector and 
$\delta V_m$ is the volume element for the $m$-th component.
The normalized quantity or vector for the $j$-th Lyapunov vector at time
$t$ is defined as 
	\begin{eqnarray}
	 \hat{\vg{\xi}}_j &=& \frac{\vg{\xi}_j(t)}{\|\vg{\xi}_j(t)\|}.
	\end{eqnarray}
Note that the first Lyapunov vector and the first Lyapunov exponent,
which we calculate here, 
do not depend on the definition of the norm.
If $t$ is large, we can rewrite the $j$-th Lyapunov vector as
	\begin{eqnarray}
	 \vg{\xi}_j(t) &=& \hat{\vg{\xi}}_j(t)\|\vg{\xi}_j(t)\| \to \hat{\vg{\xi}}_j(t)
	  e^{\lambda_j t}.
	\end{eqnarray}

Care should be taken in the numerical computation of the Lyapunov vectors
to avoid digit overflow due to their exponential growth.
Thus, these
were calculated from the results of an integration of the tangent linear
model with periodic normalizations 
starting from an arbitrary initial perturbation \citep{benettin1980lyapunov}.

\begin{figure}
\includegraphics[height=0.9\linewidth,angle=-90]{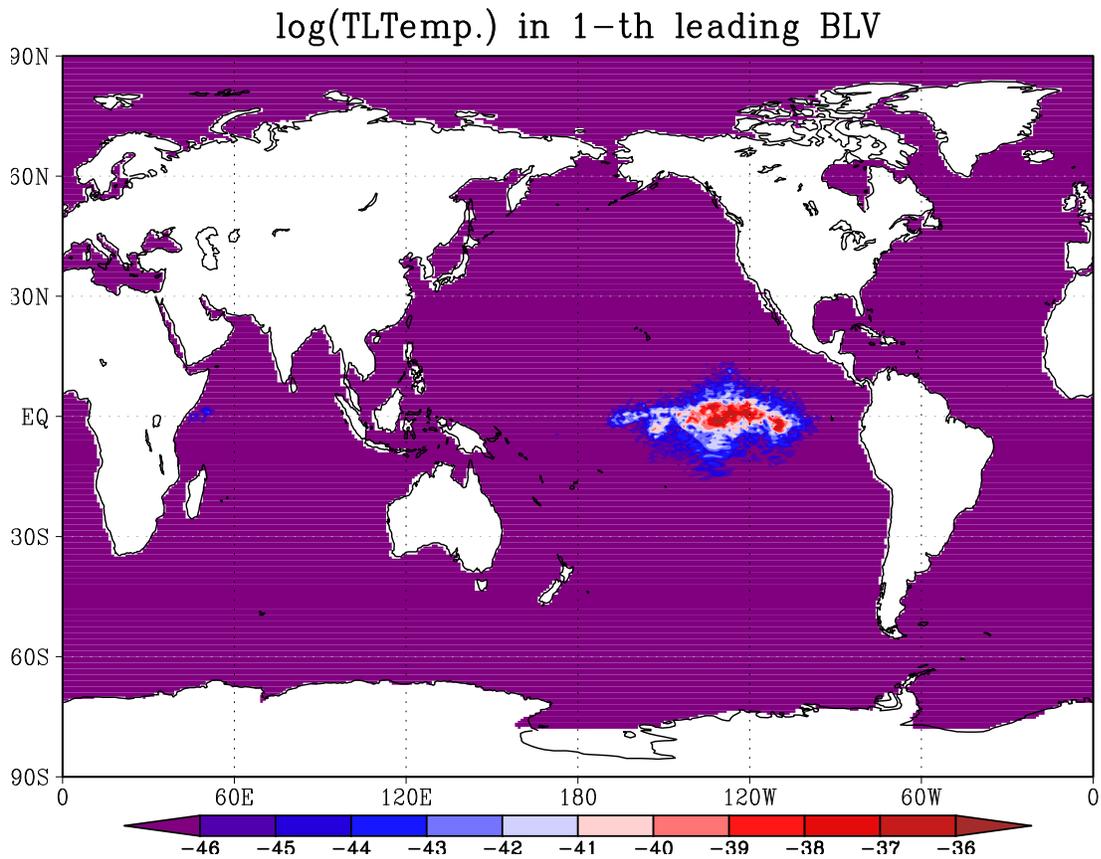}
\caption{The first  approximate backward Lyapunov vector for case 1, 
the original model, indicated with $\log{(|\delta \mathrm{SST}|)}$. 
The quantity is dimensionless because of normalization.
 The corresponding Lyapunov exponent is $1.51 \, \mathrm{day^{-1}}$.} 
\label{f2}
\end{figure}

Taking into account the highly localized feature of Lyapunov vectors
\citep{pazo2008structure}, it is convenient to characterize the Lyapunov vector by the
logarithmic quantity $\log{|\hat{\vg{\xi}}_j(t)|}$, where $|\cdot|$ denotes
the componentwise absolute value,
and by the corresponding Lyapunov exponent $\lambda_j$. 
In this case study, the spatial pattern of $\log{|\hat{\vg{\xi}}_j(t)|}$ is
well represented by the one for sea surface temperature (SST) or for subsurface water temperature.
Figure \ref{f2} shows
the first backward Lyapunov vector for case 1, whose values are shown 
as the normalized quantities of $\log{(|\delta \mathrm{SST}|)}$ 
(i.e., the logarithmic change in SST).
The time change of the vector norm shows apparent growth
corresponding to the Lyapunov exponent of $1.51 \, \mathrm{day^{-1}}$
at a location near the oceanic tropical instability waves,
suggesting the influence of 
sensitivity from the mixed-layer dynamics \citep{philander1986properties}. 
Sensitivity at this short time scale is hardly applicable to state estimation 
using an assimilation window with a seasonal-to-interdecadal time scale.

\begin{figure}
\includegraphics[height=0.9\linewidth,angle=-90]{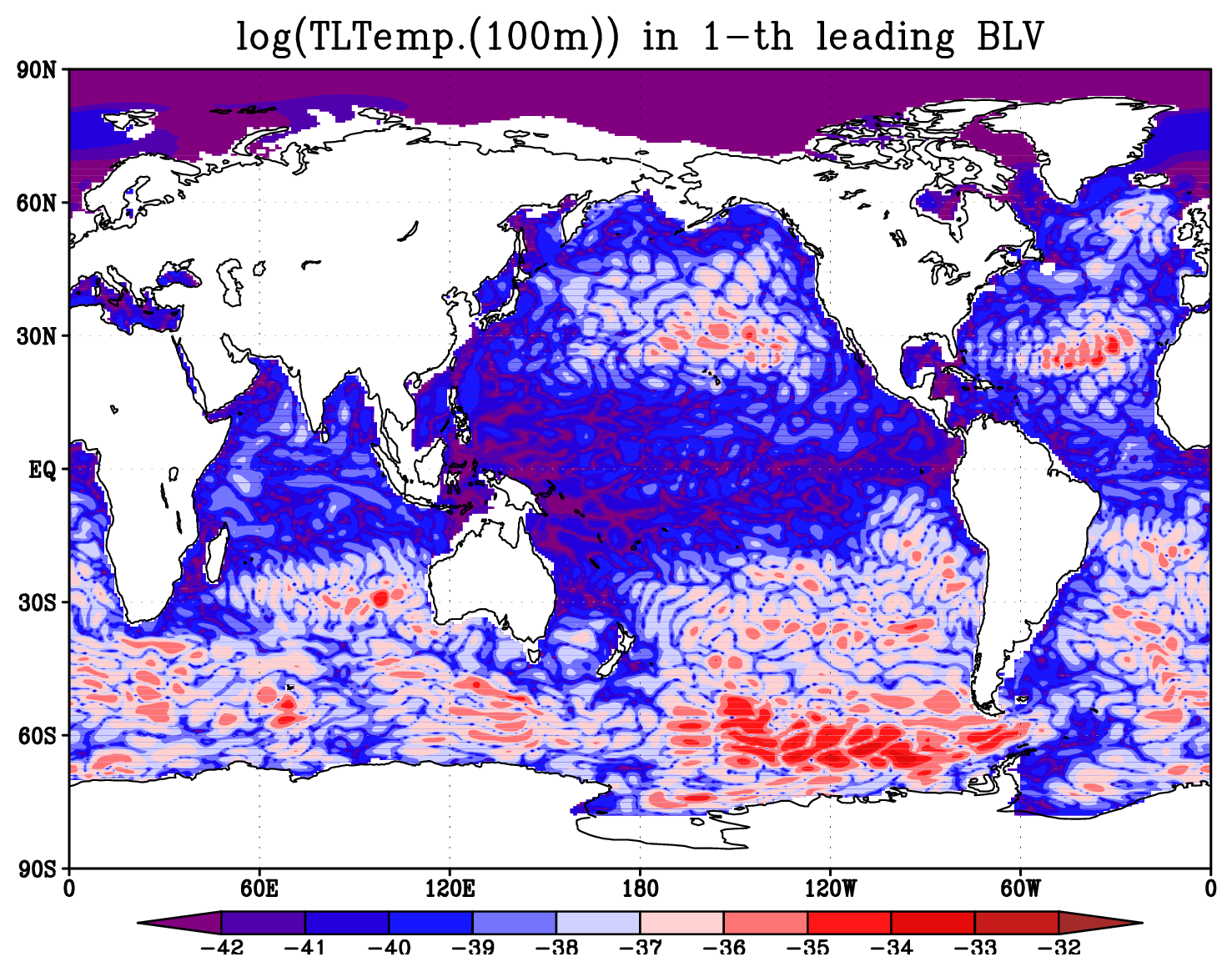}
\caption{The first approximate backward Lyapunov vector for case 2, with
a prescribed vertical diffusion coefficient, indicated with $\log{(|\delta
 \mathrm{T}(z=100 \, \mathrm{m})|)}$. 
The quantity is dimensionless because of normalization.
The corresponding Lyapunov exponent is $-0.001 \, \mathrm{day^{-1}}$.}
\label{f3}
\end{figure}

Figure \ref{f3} shows the first backward Lyapunov vector for
case 2, whose values are shown  as the normalized quantities for
$\log{(|\delta \mathrm{\mathrm{T}(z=100 \, \mathrm{m})}|)}$
(i.e., the logarithmic temperature change at the 100-m isobath). 
The corresponding Lyapunov
 exponent is $-0.001 \, \mathrm{day^{-1}}$, which
means that temporal evolution during linear model integration is
stable and that this method is relevant to seasonal-to-interdecadal
scale climate research.
The most persistent signal at the 100-m isobath appears in the area of the
      Antarctic Circumpolar Current,
and the rapidly growing signals seen in case 1 
disappear under the prescribed vertical diffusion coefficient of case 2.
Note that the short-wavelength spatial structure in the approximate Lyapunov vector
arises partly from the randomness of the initial perturbation 
because of the relatively short integration period (one year) 
of the tangent linear model.

\begin{figure}
\includegraphics[height=0.9\linewidth,angle=-90]{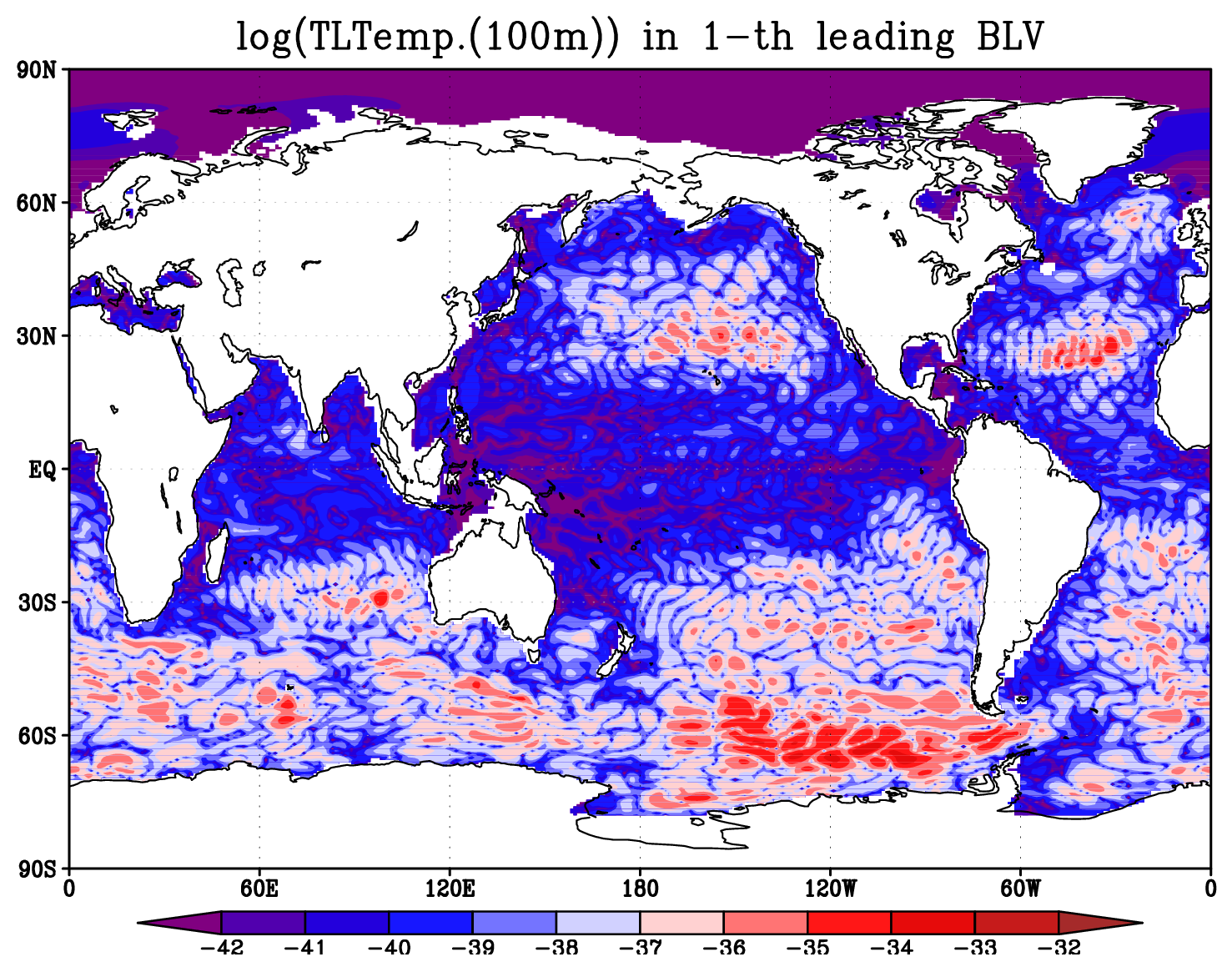}
\caption{The first approximate backward Lyapunov vector for case 3,
 with a partially
 prescribed vertical diffusion coefficient, indicated with
 $\log{(|\delta \mathrm{T}(z=100 \, \mathrm{m})|)}$. 
The quantity is dimensionless because of normalization.
The corresponding Lyapunov exponent is $-0.001 \, \mathrm{day^{-1}}$. 
}
\label{f4}
\end{figure}

Figure \ref{f4} shows 
the first backward Lyapunov vector for case 3,
whose values are also shown as the normalized quantities for $\log{(|\delta \mathrm{T}(z=100\, \mathrm{m})|)}$, 
with the corresponding Lyapunov
 exponent of $-0.001 \, \mathrm{day^{-1}}$. 
The vector exhibits almost the same structure and growth rate as case 2.
The rapidly growing signals in case 1 also disappear 
under the partially prescribed vertical diffusion coefficients of case 3.
The slight difference between cases 2 and 3 is probably caused mainly
by the numerical truncation procedure, and the two procedures have basically identical results. 

This comparison demonstrates that 
prescription of the vertical diffusion coefficient by a master equation,
either totally ($\epsilon=1$) or partially ($\epsilon=0.75$),
improves the stability of the slave system 
(Eq.\,(\ref{vdiff_sens2})).
Because 
the coupling term between the master and slave systems
determines the extent to which the vertical diffusion coefficients in
the OGCM system are prescribed,
our result helps constraining the coupling strength necessary
for regularization to function properly. 
Regularization settings that satisfy the condition $0.75\leq \epsilon \leq 1$
should at least exhibit linearly stable behavior,
and they will therefore be applicable to derive sensitivities in estimating long-term ocean states.

The result that stability is attained within some range of $\epsilon$ 
is consistent with 
our qualitative picture of how the regularization works,
although 
the quantitative argument on 
the space-time error growth in conjunction with the KPZ equation 
still remains an issue for the future.
At least it supports the interpretation that 
the regularization for vertical mixing schemes
can be explained in the framework
of synchronizing master and slave systems.%

\section{Discussion and conclusion}
We have clarified the underlying structure behind the
simplification of tangent linear and adjoint models commonly used in 4D-Var
data assimilation in conjunction with a long assimilation window of
seasonal-to-interdecadal time scale.
The concept of coupled master-slave systems \citep[e.g.,][]{pyragas1993predictable}
provides a unified interpretation
of regularization techniques 
without introducing any
ad-hoc linear models beyond adding a coupling term to the original equation.
This framework enables us to focus on stable 
temporal developments in linear models,
which are an essential component of 4D-Var data assimilation
in climate research.
The increase in stability
 helps us extracting
slowly varying perturbations relevant to climate
variations out of the many possible fundamental solutions of the linearized equation.
We demonstrated the effect of this procedure numerically with an OGCM
by changing the linearization of the functional dependency of the vertical diffusion coefficient.
The stability of the system (Eq.\,(\ref{vdiff_sens2}))
is improved when the coefficient is
partially or totally driven by a master system.
 
The 4D-Var optimization problem-setting and solution procedures 
for a practical application using this
type of master-slave setting are 
formulated as the regularized 4D-Var method.
%
Regularized 
4D-Var 
in the seasonal-to-interdecadal time scale 
can be
     viewed as an inverse problem of finding the optimal initial condition for
      the synchronized manifold of a master-slave system 
subject to observational constraints.
As the methodology is based on the stability of the synchronization
manifold under the condition
that the slave system is  selectively attracted
to the master system along unstable directions,
the design of the coupling term between the two systems is crucial for
setting up the problem
      for estimation of the long-term ocean state.
Our investigation of the existing applications of the coupling strategy
has found that 
antisymmetric coupling functions that 
selectively damp the growth of unstable modes,
by enhancing horizontal mixing in linear models
or by suppressing sensitivities of vertical mixing schemes,
are able to stabilize the slave system, but they may deform the stable
modes at the same time.
However, the contamination of the stable subspace can be reduced, 
at least in the case of enhanced horizontal mixing, by adjusting the coupling strength.
A similar interpretation is applicable to the case of 
a multi-incremental setting
\citep{courtier1994strategy,tremolet2007incremental}
(for details, see Appendix E).

The master-slave methodology is a widely applicable regularization
for climate research and long-term prediction activities,
among many other effective regularization techniques such as
the introduction of restoring terms with direct observational constraint
\citep{abarbanel2010data},
the use of a time-distributed background error \citep{cullen2010demonstration},
and the use of reduced space spanned by leading vectors derived from
empirical orthogonal function 
analysis \citep{hoteit2006efficiency,yaremchuk2009method}.

%
%
As mentioned in  \citet{evensen1997advanced},
weak constraint 4D-Var also enables us 
to extend the assimilation window 
longer than the predictability
by adding a model error term to the model equation as an additive noise.
It is implemented in practice
by dividing an assimilation window into many short predictable
sub-windows  \citep{tr2006accounting}, 
and thus it usually requires a more complicated system and more computational resources than
strong constraint 4D-Var does.
By treating the coupling intensity as an object to be minimized,
the coupling term between master and slave systems can be regarded as a kind of model error in a
weak constraint 4D-Var, 
but it should be in the form of a multiplicative noise \citep{hansen2007stochastic} 
because it is dependent on the model state of the moment. 
A weak constraint 4D-Var including additive and multiplicative noises
may offer a more general framework for climate-scale data assimilation.
Meanwhile, there still are many issues to be solved
regarding its formulation, implementation, and computational efficiency
in conjunction with regularization.%
%

\begin{acknowledgment}
We gratefully acknowledge 
many helpful comments and suggestions from several anonymous reviewers.
This study was supported in part by
the Japan Society for Promotion of Science [KAKENHI, Grant-in-aid 
for Young Scientists (B) 11024975] and the Ministry of 
Education, Culture, Sports, Science and Technology (MEXT), Japan
[Research Program on Climate Change Adaptation (RECCA) 10101028].
%
The numerical calculations were carried out on the Earth Simulator of
the Japan Agency for Marine-Earth Science and Technology (JAMSTEC). 
\end{acknowledgment}


\ifthenelse{\boolean{dc}}
{}
{\clearpage}
\begin{appendix}[A]
\section*{\begin{center}
The regularization of \citet{abarbanel2010data}
	  \end{center}}
The data assimilation problem solved by \citet{abarbanel2010data} can be
written 
as follows.\\
minimize
       \begin{eqnarray}
\mathcal{J}(\vc{x}(t_{1:N}),\mt{C}(t_{1:N}))&=& \frac{1}{2}
	 \sum_{n=1}^N 
	 \left.\left( 
		\mathrm{tr}(\mt{C}(t)^T\mt{C}(t))
		+
		(\vc{x}-\vc{y})^T
		\mt{R}^{-1}
		(\vc{x}-\vc{y})
		       \right)\right|_{t=t_n},\label{cost_abarbanel}\\
 \mt{R}^{-1}  &\equiv&
  \begin{bmatrix}
   \mt{I}_{M_u}&\\
   &\mt{O}_{M-M_u}
  \end{bmatrix},
       \end{eqnarray}
subject to 
       \begin{eqnarray}
	\dot{\vc{y}} &=& \vc{f}(\vc{y}),\label{master_Ab}\\
	\dot{\vc{x}} &=& \vc{f}(\vc{x})+
	 \vc{g}(\vc{x},\vc{y}),\label{slave_Ab}\\ 
 \vc{g}(\vc{x},\vc{y})  &\equiv&
  (\mt{D}_1 \mt{g})
  (\vc{x}-\vc{y})  =  -
  \begin{bmatrix}
   \mt{C}(t) &\\
   &\mt{O}_{M-M_u}
  \end{bmatrix}
  (\vc{x}-\vc{y}),\label{coupling_Ab}
\end{eqnarray}
where 
$\vc{y}$ is the state vector of the master system,
$\vc{x}$ is the state vector of the slave system,
$\dot{\vc{x}}$ denotes the time derivative of $\vc{x}$,
$\mt{D}_1 \mt{g}$ is the partial derivative of $\vc{g}$
with respect to the first variable,
$\mt{C}(t) \ge 0,
\dim{(\vc{x})}= M\times 1,
\dim{(\vc{y})}= M\times 1,
\dim{(\mt{C})}= M_u \times M_u$, $M$ is the dimensions of the state space,
and $M_u$ is the dimensions of
the unstable subspace
spanned by the unit vectors
$
\{ \vc{e}_i \}_{i=1,2,\cdots M_u}.
$
Obviously, the linear coupling function $\vc{g}$ is antisymmetric 
$
\vc{g}(\vc{y},\vc{x})=-\vc{g}(\vc{x},\vc{y}),
$
and the image of $\mt{D}_1 \mt{g}$ always lies in the unstable subspace.
With an appropriate choice of $\mt{C}$, the time evolution of the model described by Eq.\,(\ref{slave_Ab})
is stabilized by the constraint of observation $\vc{y}$. 
In other words, the variational equation,
\begin{eqnarray}
\dot{\delta \vc{x}} &=& \left(
\mt{D}\mt{f}
+
\mt{D}_1 \mt{g}
\right)
\delta \vc{x},\\
\mt{D}\mt{f}
&\equiv&
\begin{bmatrix}
\frac{\partial f_1}{\partial x_1}&
\frac{\partial f_1}{\partial x_2}&
\cdots&
\frac{\partial f_1}{\partial x_M}\\
\vdots&\vdots&&\vdots\\
\frac{\partial f_M}{\partial x_1}&
\frac{\partial f_M}{\partial x_2}&
\cdots&
\frac{\partial f_M}{\partial x_M}
\end{bmatrix}
\end{eqnarray}
exhibits all non-negative Lyapunov exponents
and realizes the synchronization of systems $\vc{x}$ and
$\vc{y}$,
which makes the constraint by the coupling term less and less important
as the state evolves, that is, 
$\mt{C}(t)\to 0$ as $t \to \infty$.
This is the situation in which \citet{abarbanel2010data} considered the observational data 
to be properly assimilated into the model.
The cost function of Eq.\,(\ref{cost_abarbanel}) is composed of an observational error term 
and a penalty term indicating the restoring strength.
When the synchronization $\vc{x}=\vc{y}$ is achieved after some transitional period,  
the cost converges to zero under ideal conditions.
\citet{abarbanel2010data} solved the problem numerically  by a ``direct method''
provided in the SNOPT software package \citep[e.g.,][]{barclay1998sqp}.
\label{app_abar}
\end{appendix}

\ifthenelse{\boolean{dc}}
{}
{\clearpage}
\begin{appendix}[B]
\section*{\begin{center}Decoupling the sensitivities
	  \end{center}}
To Eqs.\,(\ref{4dvar2}) and (\ref{4dvar3}),
we apply a change of variables from
$(\vc{x},\vc{y})$ to
$(\vc{v},\vc{y}')=(\vc{x}-\vc{y},\vc{y})$.
Substituting $(\vc{x},\vc{y})=(\vc{y}'+\vc{v},\vc{y}')$,
 we get
       \begin{eqnarray}
	\dot{\vc{y}'}&=&\vc{f}(\vc{y}'), \\
	\dot{\vc{y}'}+\dot{\vc{v}}&=&\vc{f}(\vc{y}'+\vc{v})+\epsilon \vc{g}(\vc{y}'+\vc{v},\vc{y}').
       \end{eqnarray}
      The difference between these two equations is
       \begin{eqnarray}
	\dot{\vc{v}}&=&\vc{f}(\vc{y}'+\vc{v})-\vc{f}(\vc{y}')+\epsilon \vc{g}(\vc{y}'+\vc{v},\vc{y}').
       \end{eqnarray}
      Taking the first-order derivative of this equation, we get
       \begin{eqnarray}
	\dot{\delta \vc{v}}&=&\mt{Df} \delta \vc{v}
	 + \epsilon \mt{D}_1 \mt{g} (\delta \vc{y}'+ \delta \vc{v})
	 + \epsilon \mt{D}_2 \mt{g}  \delta \vc{y}',
       \end{eqnarray}
      where
      $\mt{Df}$ is the derivative of $\vc{f}$ 
      and $\mt{D}_i\mt{g}$ is the partial derivative of $\vc{g}$
      with respect to the $i$-th variable. 

      If we assume the state vector
      to be in the neighborhood of the synchronization manifold $\vc{v}=0$ or $\vc{x}=\vc{y}$,
      we can consider $\mt{Df}, ~\mt{D}_1 \mt{g}$, and $\mt{D}_2 \mt{g}$ to be defined on
      $(\vc{y}',\vc{y}')$ in the combined coordinate system for the
      master and slave systems, or we could also say 
      on
      $\left(\vc{y}'+\vc{v}/2,\vc{y}'+\vc{v}/2\right)=\left((\vc{x}+\vc{y})/2,(\vc{x}+\vc{y})/2\right)$.
      In that situation, 
      $\mt{D}_1 \mt{g} = -\mt{D}_2 \mt{g}$ is deduced from the antisymmetry of $\vc{g}$.
\label{app_decoupled}
\end{appendix}

\ifthenelse{\boolean{dc}}
{}
{\clearpage}
\begin{appendix}[C]
\section*{\begin{center}Solving 4D-Var problems with regularization\end{center}}
We specify here a data assimilation setting with Eqs.\,(\ref{4dvar1}),
(\ref{4dvar2}), and (\ref{4dvar3}) in a tractable time-discretized
form, and then discuss the details of the solution procedure.
A complete nonlinear 4D-Var system with a master-slave setting aims
to find
the model state $(\vg{\psi},\vg{\theta})$  at the
initial time that approximately 
minimizes the cost function
   \begin{eqnarray}
    \mathcal{J}(\vg{\psi},\vg{\theta})
&=&
\frac12
\left\|\vg{\theta}+\vg{\psi}-\vg{\theta}_b\right\|_{\mt{B}^{-1}}^2
+
\frac12
\sum_{n=1}^N \left\|\vc{H}(\vc{x}_{n})-\vc{x}_{n}^{obs}\right\|_{\mt{R}_n^{-1}}^2
,\label{4dvar1a}\\
\vc{y}_{n} &=& \vc{F}(\vc{y}_{n-1}),\quad
 \vc{y}_0=\vg{\theta},\label{4dvar2a}\\
\vc{x}_{n} &=& \vc{F}(\vc{x}_{n-1}) - \epsilon \vc{G}(\vc{F}(\vc{x}_{n-1}),\vc{F}(\vc{y}_{n-1})),\quad
 \vc{x}_0=\vg{\theta}+\vg{\psi},\label{4dvar3a}\\
\nabla_{\vg{\psi}}\mathcal{J}&=&
\mt{B}^{-1}\left(\vg{\psi}+\vg{\theta}
		-\vg{\theta}_b\right)
\nonumber \\
&&+\sum_{n=1}^N 
\left( \frac{\partial \vc{x}_n}{\partial \vg{\psi}}\right)^T 
\mt{H}_n^T \mt{R}_n^{-1}\left[ 
\mt{H}_n 
\left( \frac{\partial \vc{x}_n}{\partial \vg{\psi}} \right)
\vg{\psi}+\vc{H}(\vc{x}_{n}(\vg{\theta}))-\vc{x}_{n}^{obs}
\right] \label{4dvar4a}
   \end{eqnarray}
where 
$n$ is the time index,
$\mt{B}$ is the background error covariance matrix,
$\mt{R}$ is the observational error covariance matrix,
$\vg{\theta}$ is the initial state of the master system,
$\vg{\psi}$ is the difference between the initial states of the master
and slave systems,
$\vg{\theta}_b$ is a first guess for the initial state of the master system,
$\vc{H}$ is an observation operator,
$\vc{x}_{n}^{obs}$ is an observation at time $n$,
$\vc{F}$ is the time-stepping operator for the original system,
and $\vc{G}$ is the coupling operator that satisfies $\vc{G}(\vc{y},\vc{x})=-\vc{G}(\vc{x},\vc{y})$.
Eq.\,(\ref{4dvar4a}) shows the gradient of the cost function 
with respect to $\vg{\psi}$ that
contains the tangent linear model
$\left( \partial \vc{x}_n/\partial \vg{\psi} \right)$
 and  the adjoint model
$\left( \partial \vc{x}_n/\partial \vg{\psi} \right)^T$.

Regarding the solution procedure for
$\min_{\vg{\theta}}\left(\min_{\vg{\psi}}\mathcal{J}\right)$,
we can solve the inner part (within the parentheses) by using a least-squares problem for $\vg{\psi}$
given $\vg{\theta}$, 
because the problem is regularized on the synchronization 
manifold, as presented in the main text.
A possible strategy for the outer part is 
to update a guess for $\vg{\theta}$ to a new guess
by a successive substitution method,
$
\vg{\theta}^{new}
=\vg{\theta}+
\argmin_{\vg{\psi}}\mathcal{J}(\vg{\psi},\vg{\theta})
\equiv \vc{h}(\vg{\theta}),$
which aims to find a fixed point $\vg{\theta}_{*}=\vc{h}(\vg{\theta}_{*})$.
This procedure 
restricts
the analysis update
direction to the difference between
the master and slave systems.
Although
the convergence of $\vg{\theta}$ 
in the reduced space 
is not 
guaranteed in the strict sense as used in \citet{gratton2008approximate},
%
if we can assume  that
the long-term evolution of the state $\vc{x}$ 
is not affected much 
by the coupling correspondent $\vc{y}$,
then 
the cost function satisfies the condition
\begin{equation}
\mathcal{J}\left(
\argmin_{\vg{\psi}}\mathcal{J}\left(\vg{\psi},\vg{\theta}\right),\vg{\theta}\right)
\simeq
\mathcal{J}\left(
\vc{0},\vg{\theta}+\argmin_{\vg{\psi}}\mathcal{J}\left(\vg{\psi},\vg{\theta}\right)\right)\label{cond_psi}.
\end{equation}
Note that the slave models used in these cost functions
share a common initial state 
$
\vc{x}_0
=
\vg{\theta}
+
\argmin_{\vg{\psi}}
\mathcal{J}
\left(\vg{\psi},\vg{\theta}\right)$.

On the basis of the above assumption, the problem is 
approximately
solved by the following iterative method, 
which is a variant of the Gauss-Newton algorithm \citep[e.g.,][]{lawless2005investigation}.

\begin{enumerate}
\item Define a first guess field $\vg{\theta}^{(k)}$ at time
      $t_0$ and iteration number $k$.
 For the first iteration, $k=0$, we choose 
      $\vg{\theta}^{(0)}=\vg{\theta}_b$, the background state.
\item 
Find the linear least-square solution $\delta\vg{\psi}=\vg{\psi}^{(k)}$ of the
      incremental cost function
       \begin{eqnarray}
    \hat{\mathcal{J}}(\delta\vg{\psi},\vg{\theta}^{(k)})
&=&
\frac12
\left\|\vg{\theta}^{(k)}-\vg{\theta}_b+\delta \vg{\psi}\right\|_{\mt{B}^{-1}}^2\nonumber\\
&+&
\frac12
\sum_{n=1}^N \left\|\vc{H}\left(\vc{x}_{n}(\vc{x}_{0}=\vg{\theta}^{(k)})\right)-\vc{x}_{n}^{obs}
+\mt{H}_n\prod_{j=1}^n\left\{ (\mt{I}-\epsilon \mt{D}_1\mt{G}_j)\mt{DF}_j
			\right\}\delta \vg{\psi}
\right\|_{\mt{R}_n^{-1}}^2,
	 \end{eqnarray}
where $\mt{D}\mt{F}$ is the derivative of $\vc{F}$ 
and $\mt{D}_1\mt{G}$ is the partial derivative of $\vc{G}$ with respect to the first variable.
\label{inc_step2}
\item Update the guess field using
   \begin{equation}
    \vg{\theta}^{(k+1)} = \vg{\theta}^{(k)}+\vg{\psi}^{(k)}\label{upd}.
   \end{equation}
   \item Repeat the procedure until a given convergence criterion is
   satisfied or a certain number of iterations has been performed.
The analysis field at the initial time is then given by
      $(\vg{\psi}^a,\vg{\theta}^a)
      =(\vg{\psi}^{(K)},\vg{\theta}^{(K)}),$
where $K$ is the total number of iterations performed.
\end{enumerate}

Each iteration of this set of steps 
forms
an outer loop. 
Within each outer loop,
the minimization problem of step \ref{inc_step2} must be solved using an
iterative procedure
known as the inner loop. 
The outer loop should work because
Eqs.\,(\ref{cond_psi}) and (\ref{upd}) imply the following sequence of
cost values:
\begin{equation}
\mathcal{J}\left(
\vg{\psi}^{(k)},\vg{\theta}^{(k)}
\right)
 \simeq
\mathcal{J}\left(\vc{0},\vg{\theta}^{(k+1)}\right)
>
\mathcal{J}\left(\vg{\psi}^{(k+1)},\vg{\theta}^{(k+1)}\right)
 \simeq
\cdots ,
\end{equation}
which is expected to decrease as the iteration proceeds.

Note that in previous applications
\citep[e.g.,][]{hoteit2005treating,sugiura2008development},
the number of inner loops in this procedure has been commonly set to 1,
which is convenient because
the state vector of the slave system is identical to that of the
master system in the first iteration.
\label{app1}
\end{appendix}

\ifthenelse{\boolean{dc}}
{}
{\clearpage}
\begin{appendix}[D]
\section*{\begin{center}Regularization of advection terms
	  \end{center}}
Suppose we have a pair of systems,
\begin{eqnarray}
\dot{\vc{y}} &=& \vc{f}_0(\vc{y})+ \vc{y} \cdot \nabla \vc{y}, \label{adv_master}\\
\dot{\vc{x}} &=& \vc{f}_0(\vc{x})+ \vc{x} \cdot \nabla \vc{x} + \epsilon 
(\vc{y}-\vc{x})     \cdot  \nabla \vc{u},\label{adv_slave}
\end{eqnarray}
where $\vc{u}=(\vc{x}+\vc{y})/2$, and
the term $\vc{y} \cdot \nabla \vc{y}$ represents the two kinds of
advection terms in an OGCM, $\vg{\mu} \cdot \nabla \vg{\tau}$ in a tracer equation and 
$\vg{\mu} \cdot \nabla \vg{\mu}$ in a momentum equation,
 where $\vg{\mu}$ is velocity and
$\vg{\tau}$ is water temperature or salinity.
The variational equation with respect to $\vc{v}=\vc{x}-\vc{y}$ in the neighborhood of
$\vc{x}=\vc{y}$ is
\begin{eqnarray}
\dot{\delta \vc{v}} &=& 
(\mt{Df}_0+ \vc{u} \cdot \nabla ) \delta \vc{v}  
+(1-\epsilon) \delta \vc{v} \cdot \nabla \vc{u}. \label{adv_var}
\end{eqnarray}
If we reduce the contribution from the last term on the right-hand side
of Eq.\,(\ref{adv_var}) by setting $0 < \epsilon \le 1$,
the stability of the slave system possibly improves by a
mechanism analogous to the stabilization by the treatment to the
vertical mixing. 
In the particular case of $\epsilon=1$,
the system should become stable  
just as a linear advection-diffusion equation should
because the terms related to the spatial gradient of the forward field are 
ignored, although this extreme setting might also cause significant loss
of information
about the sensitivities.
There remain several implementational issues to be resolved before this
method can be applied to an OGCM, for example,
modifications to a conservative formulation or
an effective discretization into linearized OGCM codes.
It is worth noting that the temporal averaging technique of the forward field for adjoint
integration \citep{sugiura2008development} could be viewed as a simplified 
variant of this treatment
in the sense that the temporal averaging is expected to smooth
the spatial gradient of the forward field used in linear models.%
	  \label{app3}
\end{appendix}

\ifthenelse{\boolean{dc}}
{}
{\clearpage}
\begin{appendix}[E]
\section*{\begin{center}Interpretation for a multi-incremental setting
	  \end{center}}
Substituting
$\vc{g}(\vc{x},\vc{y})=
(\mathcal{I}-\mathcal{P})
\left(
\vc{f}(\vc{y})-\vc{f}(\vc{x})
\right)
$
into Eqs.\,(\ref{4dvar2}), (\ref{4dvar3}), and (\ref{eq:variational})
yields nonlinear and variational equations
for a multi-incremental setting,
   \begin{eqnarray}
    \dot{\vc{y}} &=& \vc{f}(\vc{y}),\label{eq:inc1}\\
    \dot{\vc{x}} &=& \vc{f}(\vc{x})
     +
     \epsilon
     (\mathcal{I}-\mathcal{P})
     \left[
      \vc{f}(\vc{y})-\vc{f}(\vc{x})
       \right]
     ,\label{eq:inc2}\\
    \dot{\delta \vc{v}}&=&
     \left[
      (1-\epsilon) \mathcal{I} + \epsilon \mathcal{P})
	\right] 
     \mt{D}\mt{f}\delta \vc{v}, \label{eq:inc3}
   \end{eqnarray}
where
$\mathcal{I}$ is the identity operator and $\mathcal{P}$  is 
a projection operator onto a coarser grid system.

The coupling term in the slave system is designed so that the null space of
$\mathcal{P}$, that is, the variation that cannot be 
resolved by the coarse-grained description, is attracted to
the master system. 
The modified sensitivity is expected to be stabilized by being blurred by the operator $\mathcal{P}$.
In particular, if $\epsilon=1$, then the sensitivities can be
traced by
the coarser-resolution linear model  $\dot{\delta \vc{v}}
=\mathcal{P}\mt{Df}\delta \vc{v}$.
To illustrate how the stabilization works, suppose we have a simple unstable
system and a projection operator
   \begin{equation}
    \mt{Df}=
     \begin{bmatrix}
      \lambda_1 & 0\\
      0 & \lambda_2
     \end{bmatrix},
     \quad
     \mathcal{P}= 
     \frac12
     \begin{bmatrix}
      1 & 1\\
      1 & 1
     \end{bmatrix},
   \end{equation}
where $\lambda_1>0,\lambda_2<0$, and $\lambda_1+\lambda_2<0$.
%
%
Then, the growth of the $\vc{v}$-system is written, 
in terms of basis vectors $\{\vg{\xi}_j\}_{j=1,2}$, as follows.
\begin{eqnarray}
\delta  \vc{v}(t) &=& \sum_{j=1}^2 \alpha_j(t) \vg{\xi}_j,\\
\begin{bmatrix}
\dot{\alpha}_1\\
\dot{\alpha}_2
\end{bmatrix}
&=&
\mt{V}
\begin{bmatrix}
\sigma_{+}&0\\
0&\sigma_{-}
\end{bmatrix}
\mt{V}^{-1}
\begin{bmatrix}
\alpha_1\\
\alpha_2
\end{bmatrix},
\end{eqnarray}
where $\mt{V}$ is a regular matrix, and
\begin{eqnarray}
\sigma_{\pm} &=& \frac12 \left( a\pm \sqrt{a^2+b} \right),\\
a&\equiv& \left(1-\frac{\epsilon}{2} \right)(\lambda_1+\lambda_2),\\
b&\equiv& -4\left(1-\epsilon \right)(\lambda_1\lambda_2).
\end{eqnarray}
The only value allowed for $\epsilon$
in terms of stability (condition \ref{cond_b} in Section 2b) turns out to be $1$,
which leads to $\sigma_{\pm}=\left(\lambda_1+\lambda_2\right)/2, 0$.
This means that the coarser
resolution version is needed for the linear model, not the mixture of
that and the original model. 
Although from this argument the contamination of stable modes might seem
destructive,
that is not necessarily the case in actual N-dimensional systems,
where the evolution matrix will be in a block diagonal form
\begin{equation}
\frac12
\diag{(\lambda_1,\lambda_2,\cdots,\lambda_N)}
\begin{bmatrix}
1&1& & & & & \\
1&1& & & & & \\
 & &1&1& & & \\
 & &1&1&\ddots & & \\
 & & &\ddots & &1&1\\
 & & & & &1&1
\end{bmatrix}.
\end{equation}
Obviously, this system represents sensitivities that have a coarse-grained growth rate $\left(\lambda_i+\lambda_{i+1}\right)/2$,
just as a coarser resolution model does.%

\citet{gebbie2006strategies}, who applied this type of regularization to
ocean data assimilation,
successfully used a coarser resolution adjoint OGCM for an eddy-permitting state estimation.%
\label{app2}
\end{appendix}

\ifthenelse{\boolean{dc}}
{}
{\clearpage}
\bibliographystyle{ametsoc}
\bibliography{./ref}

\begin{thebibliography}{42}
\providecommand{\natexlab}[1]{#1}
\providecommand{\url}[1]{\texttt{#1}}
\providecommand{\urlprefix}{URL }
\expandafter\ifx\csname urlstyle\endcsname\relax
  \providecommand{\doi}[1]{doi:\discretionary{}{}{}#1}\else
  \providecommand{\doi}{doi:\discretionary{}{}{}\begingroup
  \urlstyle{rm}\Url}\fi
\providecommand{\eprint}[2][]{\url{#2}}

\bibitem[{Abarbanel et~al.(2010)Abarbanel, Kostuk, and
  Whartenby}]{abarbanel2010data}
Abarbanel, H. D.~I., M.~Kostuk, and W.~Whartenby, 2010: Data assimilation with
  regularized nonlinear instabilities. \textit{Quart. J. Roy. Meteor. Soc.},
  \textbf{136~(648)}, 769--783.

\bibitem[{Barclay et~al.(1998)Barclay, Gill, and Rosen}]{barclay1998sqp}
Barclay, A., P.~E. Gill, and J.~B. Rosen, 1998: {SQP} methods and their
  application to numerical optimal control. \textit{International series of
  numerical mathematics}, 207--222.

\bibitem[{Benettin et~al.(1980)Benettin, Galgani, Giorgilli, and
  Strelcyn}]{benettin1980lyapunov}
Benettin, G., L.~Galgani, A.~Giorgilli, and J.~M. Strelcyn, 1980: Lyapunov
  characteristic exponents for smooth dynamical systems and for {H}amiltonian
  systems; a method for computing all of them. {P}art 1: {T}heory.
  \textit{Meccanica}, \textbf{15~(1)}, 9--20.

\bibitem[{Buizza(1994)}]{buizza1994sensitivity}
Buizza, R., 1994: Sensitivity of optimal unstable structures. \textit{Quart. J.
  Roy. Meteor. Soc.}, \textbf{120~(516)}, 429--451.

\bibitem[{Courtier et~al.(1994)Courtier, Th{\'e}paut, and
  Hollingsworth}]{courtier1994strategy}
Courtier, P., J.~N. Th{\'e}paut, and A.~Hollingsworth, 1994: A strategy for
  operational implementation of 4{D}-{V}ar, using an incremental approach.
  \textit{Quart. J. Roy. Meteor. Soc.}, \textbf{120~(519)}, 1367--1387.

\bibitem[{Cullen(2010)}]{cullen2010demonstration}
Cullen, M. J.~P., 2010: A demonstration of 4{D}-{V}ar using a time-distributed
  background term. \textit{Quart. J. Roy. Meteor. Soc.}, \textbf{136~(650)},
  1301--1315.

\bibitem[{Duane et~al.(2006)Duane, Tribbia, and Weiss}]{duane2006synchronicity}
Duane, G.~S., J.~J. Tribbia, and J.~B. Weiss, 2006: Synchronicity in predictive
  modelling: a new view of data assimilation. \textit{Nonlinear Processes in
  Geophysics}, \textbf{13~(6)}, 601--612.

\bibitem[{Evensen(1997)}]{evensen1997advanced}
Evensen, G., 1997: Advanced data assimilation for strongly nonlinear dynamics.
  \textit{Mon. Wea. Rev.}, \textbf{125~(6)}, 1342--1354.

\bibitem[{Gebbie et~al.(2006)Gebbie, Heimbach, and
  Wunsch}]{gebbie2006strategies}
Gebbie, G., P.~Heimbach, and C.~Wunsch, 2006: Strategies for nested and
  eddy-permitting state estimation. \textit{J. Geophys. Res.},
  \textbf{111~(C10)}, C10\,073.

\bibitem[{Gratton et~al.(2008)Gratton, Lawless, and
  Nichols}]{gratton2008approximate}
Gratton, S., A.~Lawless, and N.~Nichols, 2008: {Approximate Gauss-Newton
  methods for nonlinear least squares problems}. \textit{SIAM Journal on
  Optimization}, \textbf{18~(1)}, 106--132, \doi{10.1137/050624935}.

\bibitem[{Hansen and Penland(2007)}]{hansen2007stochastic}
Hansen, J.~A. and C.~Penland, 2007: On stochastic parameter estimation using
  data assimilation. \textit{Physica D: Nonlinear Phenomena}, \textbf{230~(1)},
  88--98.

\bibitem[{Hoteit et~al.(2005)Hoteit, Cornuelle, K{\"o}hl, and
  Stammer}]{hoteit2005treating}
Hoteit, I., B.~Cornuelle, A.~K{\"o}hl, and D.~Stammer, 2005: Treating strong
  adjoint sensitivities in tropical eddy-permitting variational data
  assimilation. \textit{Quart. J. Roy. Meteor. Soc.}, \textbf{131~(613)},
  3659--3682.

\bibitem[{Hoteit and K\"{o}hl(2006)}]{hoteit2006efficiency}
Hoteit, I. and A.~K\"{o}hl, 2006: Efficiency of reduced-order, time-dependent
  adjoint data assimilation approaches. \textit{Journal of Oceanography},
  \textbf{62~(4)}, 539--550.

\bibitem[{Johnson et~al.(2006)Johnson, Hoskins, and
  Nichols}]{johnson2006singular}
Johnson, C., B.~J. Hoskins, and N.~K. Nichols, 2006: A singular vector
  perspective of {4D-Var}: {F}iltering and interpolation. \textit{Quart. J.
  Roy. Meteor. Soc.}, \textbf{131~(605)}, 1--19.

\bibitem[{Kapral et~al.(1994)Kapral, Livi, Oppo, and
  Politi}]{kapral1994dynamics}
Kapral, R., R.~Livi, G.~Oppo, and A.~Politi, 1994: Dynamics of complex
  interfaces. \textit{Physical Review E}, \textbf{49~(3)}, 2009.

\bibitem[{Kardar et~al.(1986)Kardar, Parisi, and Zhang}]{kardar1986dynamic}
Kardar, M., G.~Parisi, and Y.~Zhang, 1986: Dynamic scaling of growing
  interfaces. \textit{Physical Review Letters}, \textbf{56~(9)}, 889--892.

\bibitem[{K{\"o}hl and Stammer(2008)}]{kohl2008decadal}
K{\"o}hl, A. and D.~Stammer, 2008: Decadal sea level changes in the 50-year
  gecco ocean synthesis. \textit{J. Climate}, \textbf{21~(9)}, 1876--1890,
  \doi{10.1175/2007JCLI2081.1}.

\bibitem[{K{\"o}hl and Willebrand(2008)}]{kohl2008adjoint}
K{\"o}hl, A. and J.~Willebrand, 2008: An adjoint method for the assimilation of
  statistical characteristics into eddy-resolving ocean models. \textit{Tellus
  A}, \textbf{54~(4)}, 406--425.

\bibitem[{Lawless et~al.(2005)Lawless, Gratton, and
  Nichols}]{lawless2005investigation}
Lawless, A.~S., S.~Gratton, and N.~K. Nichols, 2005: An investigation of
  incremental 4{D}-{V}ar using non-tangent linear models. \textit{Quart. J.
  Roy. Meteor. Soc.}, \textbf{131~(606)}, 459--476.

\bibitem[{Lea et~al.(2002)Lea, Haine, Allen, and Hansen}]{lea2002sensitivity}
Lea, D.~J., T.~W.~N. Haine, M.~R. Allen, and J.~A. Hansen, 2002: Sensitivity
  analysis of the climate of a chaotic ocean circulation model. \textit{Quart.
  J. Roy. Meteor. Soc.}, \textbf{128~(586)}, 2587--2605.

\bibitem[{Legras and Vautard(1996)}]{legras1996guide}
Legras, B. and R.~Vautard, 1996: A guide to {L}iapunov vectors.
  \textit{Proceedings 1995 ECMWF Seminar on Predictability}, Vol.~1, 143--156.

\bibitem[{Luenberger(1964)}]{luenberger1964observing}
Luenberger, D.~G., 1964: Observing the state of a linear system. \textit{IEEE
  Transactions on Military Electronics}, \textbf{8~(2)}, 74--80.

\bibitem[{Masuda et~al.(2010)}]{masuda2010simulated}
Masuda, S., et~al., 2010: Simulated rapid warming of abyssal north pacific
  waters. \textit{Science}, \textbf{329~(5989)}, 319--322.

\bibitem[{Mazloff et~al.(2010)Mazloff, Heimbach, and Wunsch}]{mazloff2010eddy}
Mazloff, M.~R., P.~Heimbach, and C.~Wunsch, 2010: {An eddy-permitting Southern
  Ocean state estimate}. \textit{J. Phys. Oceanogr.}, \textbf{40~(5)},
  880--899.

\bibitem[{Noh and Kim(1999)}]{noh1999simulations}
Noh, Y. and H.~J. Kim, 1999: Simulations of temperature and turbulence
  structure of the oceanic boundary layer with the improved near-surface
  process. \textit{J. Geophys. Res.}, \textbf{104}, 15.

\bibitem[{Paz{\'o} et~al.(2008)Paz{\'o}, Szendro, L{\'o}pez, and
  Rodr{\'\i}guez}]{pazo2008structure}
Paz{\'o}, D., I.~G. Szendro, J.~L{\'o}pez, and M.~A. Rodr{\'\i}guez, 2008:
  Structure of characteristic {L}yapunov vectors in spatiotemporal chaos.
  \textit{Physical Review E}, \textbf{78~(1)}, 016\,209.

\bibitem[{Pecora and Carroll(1990)}]{pecora1990synchronization}
Pecora, L.~M. and T.~L. Carroll, 1990: Synchronization in chaotic systems.
  \textit{Physical Review Letters}, \textbf{64~(8)}, 821--824.

\bibitem[{Philander et~al.(1986)Philander, Hurlin, and
  Pacanowski}]{philander1986properties}
Philander, S. G.~H., W.~J. Hurlin, and R.~C. Pacanowski, 1986: Properties of
  long equatorial waves in models of the seasonal cycle in the tropica
  {A}tlantic and {P}acific {O}ceans. \textit{J. Geophys. Res.},
  \textbf{91~(C12)}, 14\,207--14\,211.

\bibitem[{Primo et~al.(2007)Primo, Szendro, Rodr{\'\i}guez, and
  Guti{\'e}rrez}]{primo2007error}
Primo, C., I.~Szendro, M.~Rodr{\'\i}guez, and J.~Guti{\'e}rrez, 2007: Error
  growth patterns in systems with spatial chaos: from coupled map lattices to
  global weather models. \textit{Physical Review Letters}, \textbf{98~(10)},
  108\,501.

\bibitem[{Pyragas(1993)}]{pyragas1993predictable}
Pyragas, K., 1993: Predictable chaos in slightly perturbed unpredictable
  chaotic systems. \textit{Physics Letters A}, \textbf{181~(3)}, 203--210.

\bibitem[{So et~al.(1994)So, Ott, and Dayawansa}]{so1994observing}
So, P., E.~Ott, and W.~P. Dayawansa, 1994: Observing chaos: {D}educing and
  tracking the state of a chaotic system from limited observation.
  \textit{Physical Review E}, \textbf{49~(4)}, 2650--2660.

\bibitem[{Stammer et~al.(2002)}]{stammer2002}
Stammer, D., et~al., 2002: Global ocean circulation during 1992--1997,
  estimated from ocean observations and a general circulation model. \textit{J.
  Geophys. Res.}, \textbf{107}, 3118, \doi{10.1029/2001JC000888}.

\bibitem[{Sugiura et~al.(2008)Sugiura, Awaji, Masuda, Mochizuki, Toyoda,
  Miyama, Igarashi, and Ishikawa}]{sugiura2008development}
Sugiura, N., T.~Awaji, S.~Masuda, T.~Mochizuki, T.~Toyoda, T.~Miyama,
  H.~Igarashi, and Y.~Ishikawa, 2008: Development of a four-dimensional
  variational coupled data assimilation system for enhanced analysis and
  prediction of seasonal to interannual climate variations. \textit{J. Geophys.
  Res.}, \textbf{113~(C10)}, C10\,017.

\bibitem[{Tr\'{e}molet(2006)}]{tr2006accounting}
Tr\'{e}molet, Y., 2006: Accounting for an imperfect model in {4D-Var}.
  \textit{Quart. J. Roy. Meteor. Soc.}, \textbf{132~(621)}, 2483--2504.

\bibitem[{Tr{\'e}molet(2007)}]{tremolet2007incremental}
Tr{\'e}molet, Y., 2007: Incremental {4D-Var} convergence study. \textit{Tellus
  A}, \textbf{59~(5)}, 706--718.

\bibitem[{Tsujino et~al.(2011)Tsujino, Hirabara, Nakano, Yasuda, Motoi, and
  Yamanaka}]{mricom2}
Tsujino, H., M.~Hirabara, H.~Nakano, T.~Yasuda, T.~Motoi, and G.~Yamanaka,
  2011: Simulating present climate of the global ocean-ice system using the
  {Meteorological Research Institute Community Ocean Model (MRI.COM)}:
  simulation characteristics and variability in the {P}acific sector.
  \textit{Journal of Oceanography}, \textbf{67}, 449--479.

\bibitem[{Tsujino et~al.(2010)Tsujino, Motoi, Ishikawa, Hirabara, Nakano,
  Yamanaka, and Yasuda}]{mricom1}
Tsujino, H., T.~Motoi, I.~Ishikawa, M.~Hirabara, H.~Nakano, G.~Yamanaka, and
  T.~Yasuda, 2010: {Meteorological Research Institute Community Ocean Model
  Version 3 (MRI.COM3) Manual}. Technical Report of Meteorological Research
  Institute~59, Meteorological Research Institute, Tsukuba, Japan. 241pp.

\bibitem[{Wunsch and Heimbach(2007)}]{wunsch2007practical}
Wunsch, C. and P.~Heimbach, 2007: Practical global oceanic state estimation.
  \textit{Physica D: Nonlinear Phenomena}, \textbf{230~(1)}, 197--208.

\bibitem[{Yang et~al.(2006)}]{yang2006data}
Yang, S.~C., et~al., 2006: {D}ata {A}ssimilation as {S}ynchronization of
  {T}ruth and {M}odel: {E}xperiments with the {T}hree-{V}ariable {L}orenz
  {S}ystem. \textit{J. Atmos. Sci.}, \textbf{63~(9)}, 2340--2354.

\bibitem[{Yaremchuk et~al.(2009)Yaremchuk, Nechaev, and
  Panteleev}]{yaremchuk2009method}
Yaremchuk, M., D.~Nechaev, and G.~Panteleev, 2009: A {M}ethod of {S}uccessive
  {C}orrections of the {C}ontrol {S}ubspace in the {R}educed-{O}rder
  {V}ariational {D}ata {A}ssimilation. \textit{Mon. Wea. Rev.},
  \textbf{137~(9)}, 2966--2978.

\bibitem[{Zhu and Kamachi(2000)}]{zhu2000role}
Zhu, J. and M.~Kamachi, 2000: The role of time step size in numerical stability
  of tangent linear models. \textit{Mon. Wea. Rev.}, \textbf{128~(5)},
  1562--1572.

\bibitem[{Zhu et~al.(2002)Zhu, Kamachi, and Wang}]{Zhu2002}
Zhu, J., M.~Kamachi, and D.~Wang, 2002: Estimation of air-sea heat flux from
  ocean measurements: {A}n ill-posed problem. \textit{J. Geophys. Res.},
  \textbf{107~(C10)}, 3159, \doi{10.1029/2001JC000995}.

\end{thebibliography}

\end{document}